\documentclass[a4paper,11pt]{article}
\pdfoutput=1 

\usepackage{jinstpub} 

\usepackage{graphicx}
\usepackage{latexsym}
\usepackage{subfig}
\usepackage{amssymb}

\title{The Extreme Energy Events experiment: an overview of the telescopes performance.}

\author[1,2]{M.Abbrescia,}
\author[1,3]{C.Avanzini,}  
\author[1,4]{L.Baldini Ferroli,} 
\author[1,3]{L.Baldini,} 
\author[1,3]{G.Batignani,} 
\author[1,17]{M.Battaglieri,}  
\author[1,5]{S.Boi,} 
\author[1,15,5]{E.Bossini,} 
\author[1,6]{F.Carnesecchi,}
\author[1,7]{A.Chiavassa,} 
\author[1,8]{C.Cicalo,} 
\author[1,6]{L.Cifarelli,} 
\author[1]{F.Coccetti,} 
\author[1,9]{E.Coccia,}  
\author[1,10]{A.Corvaglia,}  
\author[1,11]{D.De Gruttola,}  
\author[1,11]{S.De Pasquale,} 
\author[1,4]{F.L.Fabbri,} 
\author[16]{V.Frolov,} 
\author[1,7]{L.Galante,}
\author[1,7]{P.Galeotti,}
\author[1,6]{M.Garbini,} 
\author[1,17]{G.Gemme,} 
\author[1,7]{I.Gnesi,}
\author[1]{S.Grazzi,}
\author[1,12]{C.Gustavino,} 
\author[1,6,15]{D.Hatzifotiadou,} 
\author[1,18]{P.La Rocca,}   
\author[1,19]{G.Mandaglio,}  
\author[14]{O.Maragoto Rodriguez,}
\author[13]{G.Maron,}  
\author[1,20]{M.N.Mazziotta,}
\author[1,4]{S.Miozzi,} 
\author[1,6]{R.Nania,} 
\author[1,6]{F.Noferini,} 
\author[1,21]{F.Nozzoli,} 
\author[1,6]{F.Palmonari,} 
\author[1,10]{M.Panareo,} 
\author[1,10]{M.P.Panetta,} 
\author[1,5]{R.Paoletti,} 
\author[14]{W.Park,} 
\author[13]{C.Pellegrino,} 
\author[1,17]{L.Perasso,} 
\author[1,3]{F.Pilo,} 
\author[1,7]{G.Piragino,} 
\author[1,4]{S.Pisano,}
\author[1,18]{F.Riggi,}
\author[1]{G.C.Righini,}  
\author[1,11]{C.Ripoli,}  
\author[1,2]{M.Rizzi,}  
\author[1,6]{G.Sartorelli,}
\author[1,6]{E.Scapparone,} 
\author[1,22]{M.Schioppa,} 
\author[1,3]{A.Scribano,} 
\author[1,6]{M.Selvi,}  
\author[1,8]{S.Serci,} 
\author[1,17]{S.Squarcia,}  
\author[1,17]{M.Taiuti,} 
\author[1,3]{G.Terreni,} 
\author[1,23]{A.Trifir\`{o},} 
\author[1,23]{M.Trimarchi,}  
\author[13]{M.C.Vistoli,} 
\author[1,12]{L.Votano,} 
\author[1,6,15]{M.C.S.Williams,}  
\author[1,14,15]{L.Zheng,} 
\author[1,6,15]{A.Zichichi,}  
\author[1,14,15]{R.Zuyeuski}

\affiliation[1]{Museo Storico
della Fisica e Centro Studi e Ricerche "Enrico Fermi", Roma, Italy}
\affiliation[2]{INFN and Dipartimento Interateneo di Fisica, Universit\`{a} di Bari, Bari, Italy} 
\affiliation[3]{INFN and
Dipartimento di Fisica, Universit\`{a} di Pisa, Pisa, Italy} 
\affiliation[4]{INFN, Laboratori Nazionali di Frascati, Frascati (RM), Italy}
\affiliation[5]{INFN Gruppo Collegato di Siena and Dipartimento di Fisica,
Universit\`{a} di Siena, Siena, Italy}
\affiliation[6]{INFN and Dipartimento di Fisica e Astronomia, Universit\`{a} di Bologna, Bologna,
Italy}
\affiliation[7]{INFN and Dipartimento di Fisica, Universit\`{a} di
Torino, Torino, Italy}
\affiliation[8]{INFN and Dipartimento di Fisica,
Universit\`{a} di Cagliari, Cagliari, Italy}
\affiliation[9]{INFN and Dipartimento di Fisica, Universit\`{a} di
Roma Tor Vergata, Roma, Italy}
\affiliation[10]{INFN and Dipartimento di Matematica e Fisica, Universit\`{a} del Salento, Lecce, Italy}
\affiliation[11]{INFN and
Dipartimento di Fisica, Universit\`{a} di Salerno, Salerno, Italy}
\affiliation[12]{INFN, Laboratori
Nazionali del Gran Sasso, Assergi (AQ), Italy}
\affiliation[13]{INFN CNAF, Bologna, Italy}
\affiliation[14]{ICSC World Laboratory, Geneva, Switzerland}
\affiliation[15]{CERN, Geneva, Switzerland}
\affiliation[16]{JINR Joint Institute for Nuclear Research, Dubna, Russia}
\affiliation[17]{INFN and Dipartimento di Fisica, Universit\`{a} di
Genova, Genova, Italy}
\affiliation[18]{INFN and Dipartimento di Fisica e Astronomia,
Universit\`{a} di Catania, Catania, Italy}
\affiliation[19]{INFN Sezione di Catania and Dipartimento di Scienze Chimiche, Biologiche, Farmaceutiche e Ambientali,
Universit\`{a} di Messina, Messina, Italy}
\affiliation[20]{INFN Sezione di Bari, Bari, Italy}
\affiliation[21]{INFN and ASI Science Data Center, Roma, Italy}
\affiliation[22]{INFN and Dipartimento di Fisica,
Universit\`{a} della Calabria, Cosenza, Italy}
\affiliation[23]{INFN Sezione di Catania and Dipartimento di Scienze Matematiche e Informatiche, Scienze Fisiche e Scienze della Terra, Universit\`{a} di Messina, Messina, Italy} 

\abstract{The muon telescopes of the Extreme Energy Events (EEE) experiment are based on
Multigap Resistive Plate Chambers (MRPC). 
The EEE network is composed, so far, of 53 telescopes, each made of
three MRPC detectors; it is organized in clusters and single telescope stations 
distributed all over the Italian territory and installed in High Schools, covering an area larger than $3\times10^{5}$ km$^{2}$.
The study of Extensive Air Showers (EAS), that is one of the goal of the project, requires excellent performance in terms of time and spatial resolution, efficiency, tracking capability and long term stability.
The data from two recent coordinated data taking periods, named Run 2 and Run 3, have been used to measure these quantities and the results are here reported,
together with a comparison with expectations and with the results from a beam test performed in 2006 at CERN. }


\begin{document}
\maketitle
\flushbottom

\section{Introduction}
\label{sec:intro}

The Extreme Energy Events (EEE) experiment \cite{eee} is a project by Centro Fermi (Museo Storico della Fisica e Centro Studi e Ricerche "Enrico Fermi") \cite{CFsite}, in collaboration with INFN (Istituto Nazionale di Fisica Nucleare), CERN (European Council for Nuclear Research) and MIUR (the Italian Ministry of Education, University and Research).
EEE is designed to study Cosmic Rays (CR) and CR-related phenomena, via a synchronous sparse network of 53 tracking detectors, deployed over an area covering more than 10 degrees in latitude and 11 in longitude, corresponding to more than 3 $\times$ $10^{5}$ km$^{2}$. The observatory is being upgraded regularly: 6 new stations have been  installed within the end of 2017, leading to a 10$\%$ increase in the number of telescopes.\\
The EEE network is composed both by clusters and stand-alone stations; the result is a sparse network where each detection site is located at distances ranging from 15 m to several km from the one nearby. Each station is made of three Multigap Resistive Plate Chambers (MRPC), a CR dedicated version of the detector 
successfully used for Time Of Flight (TOF) systems and tracking detectors in high energy physics experiments at colliders (examples are  the TOF system \cite{tof1} of the ALICE experiment \cite{alice} at LHC and of the STAR experiment at RHIC \cite{star}). 
Data collected by each station are sent to the CNAF center \cite{cnaf}, the computing facility of the INFN, where they are stored, reconstructed and made available for analysis. The experiment has completed its Run 3 in June 2017; the whole data set collected since fall 2014 has already exceeded 50 billions of muon track candidates.
\\The topology of the EEE network allows to measure  time-correlated events at distances never addressed before. Telescopes placed in the same city can detect individual EAS \cite{coinc1}, whereas telescopes located hundreds of kilometers apart can, in principle, detect the coincidence between two different correlated air showers, for which a few interesting events were found.\\
The EEE network can also address the local
properties of the CR flux and its space weather-correlated features \cite{forbushDec2,forbushDec3}, CR flux anisotropies in the sub-TeV energy region \cite{anisotropies} and also phenomena related to the upward-going particle flux \cite{upward}.\\
The EEE project has also a strong outreach impact: 47 detectors are installed in High Schools, where students and
teachers actively participate to the data taking activities, taking care of the telescope operation and maintenance. Researchers coordinate and supervise activities, providing
support during detectors construction, installation and operation. Students and teachers are introduced through seminars, lectures and master-classes, to the scientific research community, with the opportunity of understanding how a real experiment works, from the infrastructure development to the data acquisition, analysis and publications of scientific results.\\
The EEE network commenced its operational activity in 2004 with a set of pilot sites in 7 Italian cities. In 2017 the observatory has grown up by a factor almost 8 in terms of number of telescopes. The EEE network is the largest and long-living MRPC-based system, with 53 sites instrumented and more than 12 years of data taking. The unconventional working sites, mainly school buildings with non-professional electrical lines, non-controlled environmental parameters and heterogeneous maintenance conditions, are a unique test field for checking the robustness, the ageing and the long-lasting performance of the MRPC technology for particle tracking and timing determination. 
In addition to the schools already hosting a telescope, 54 more institutes joined the project despite not being equipped with the detector; all the students contribute to the experiment by monitoring the telescopes performance and analyzing the available data.

\section{MRPC for the EEE telescopes}
\label{eee_mrpc} 
\begin{figure}[htb]
\centerline{
\includegraphics[width=0.8\columnwidth]{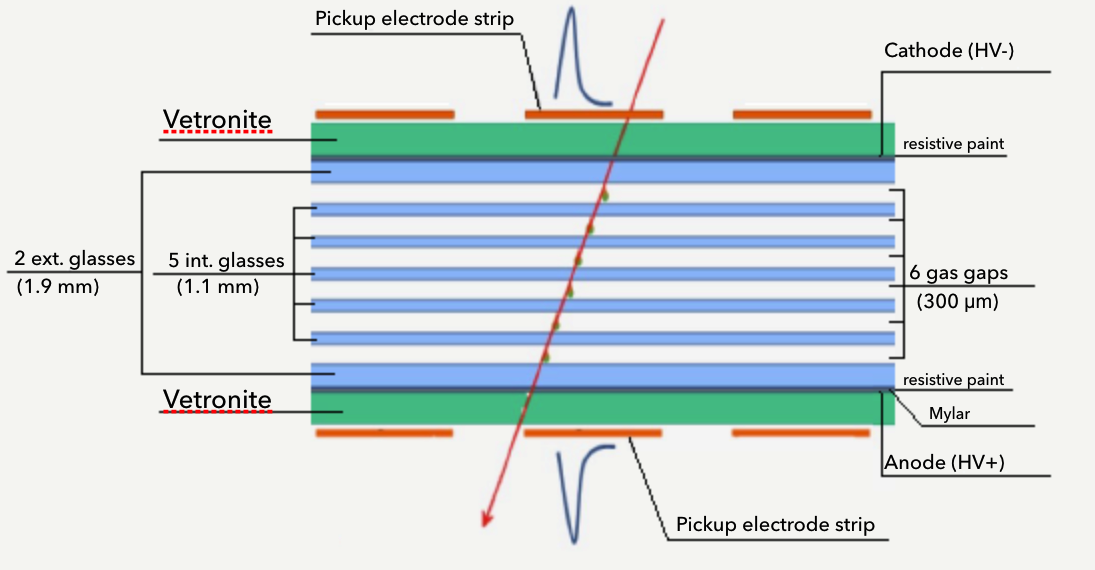}
}
\caption{\footnotesize MRPC inner structure.}
\label{mrpc}
\end{figure}

\begin{figure}[htb]
\centerline{
\includegraphics[width=0.9\columnwidth]{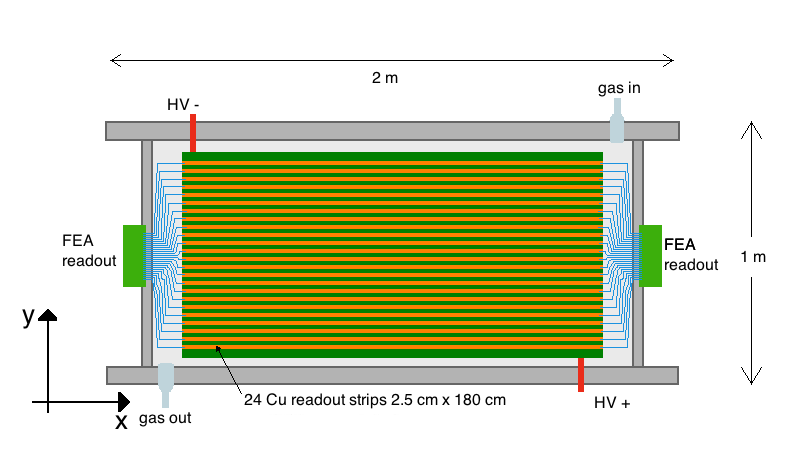}
}
\caption{\footnotesize Top view of a MRPC chamber, showing the 24 copper strips,
  the 2 FEA at the edges of the chamber and the connectors where 2 DC/DC converters providing the HV (top left in blu and bottom right in red) are plugged in.}
\label{fea_chamber}
\end{figure}

The chambers composing the EEE telescopes are MRPC detectors specifically designed for combining good tracking and timing capabilities, low construction costs and easy assembly procedures. Since the students of each participating school are directly involved in the construction of their own detectors, it is therefore important that the materials used are easy to find, safe and simple to assemble. The detector structure (Fig. \ref{mrpc}) consists  of  6  gas  gaps  obtained  by  stacking  glass  sheets, with voltage  applied  only  to  the  external  ones,  and  leaving  the  inner  ones floating. 
The cathode and the anode consist of two glasses (160 cm $\times$ 85 cm, 1.9 mm thick) treated with resistive paint (5-20 M$\Omega$/$\Box$) connected to high voltage, the space between them being divided into the six narrow gaps (300 $\mu$m) by 5 intermediate  glass sheets (158 cm $\times$ 82 cm, 1.1 mm thick); inner-glass spacing is assured through a weave made with fishing line. Anode and cathode are also in contact, on the outer surfaces, with a sheet of Mylar (175 cm $\times$ 86 cm in dimensions) stretched on a vetronite panel of equal area on whose external surface 24 copper strips are laid out (180 cm $\times$ 2.5 cm spaced by 7 mm), to collect the signals induced by particles.
Two rigid composite honeycomb panels (180 cm $\times$ 90 cm) are used to assure good mechanical stability to the whole structure, which is enclosed in a gas-tight aluminum box (220 cm $\times$ 110 cm of external dimensions, 192 cm $\times$ 92 cm inside). A schematic top view of a chamber is shown in Fig. \ref{fea_chamber}. The gas inlets and outlets, and the high voltage connectors are located at the ends of the longer sides, while the front-end (FEA) boards for the read-out of the strip signals are placed on the short sides. 
\\Chambers are filled with a gas mixture consisting of a 98/2 mixture of R134a (C$_{2}$F$_{4}$H$_{2}$) and SF$_{6}$, at a continuous flow of 2 l/h and atmospheric pressure. High voltage to the chambers, typically in the 18-20 kV range, is provided by a set of DC/DC converters, with output voltage roughly a factor 2000 with respect to the driving low voltage (LV).
Stand-alone LV power supply units, both commercial or custom engineered by the EEE Collaboration, provide the LV to the DC-DC converters. The core unit of the DC-DC converters are the EMCO Q-series, both positive and negative, with a 10 kV full scale output. The HV stability declared by the manifacturer is $\pm$10\% at full load (50 $\mu$A). The typical working voltage is 8-9 kV, thus very close to the full scale. \\

\begin{figure}[htb]
\centerline{
\includegraphics[width=0.6\columnwidth]{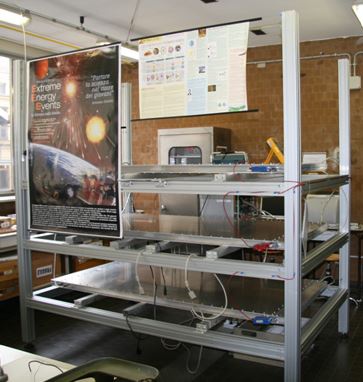}
}
\caption{\footnotesize One EEE Project telescope located at Liceo Classico Massimo D'Azeglio in Turin, composed of three MRPC chambers.}
\label{telescope}
\end{figure}

The aforementioned 24 copper strips that collect the signal, provide two-dimensional information when a cosmic muon crosses the chamber; in our reference system:

\begin{itemize}
\item the $y$ coordinate is determined by the strip on which the signal is induced;
\item the $x$ coordinate is determined by measuring the difference between the arrival time of the signal at the two ends of the strip.
\end{itemize}
FEA cards (2 for each chamber) incorporate the ultrafast and low power NINO ASIC amplifier/discriminator specifically designed for MRPC operation \cite{asic}. Three MRPC chambers assembled in a telescope are shown in Fig. \ref{telescope}.

The trigger logic consists in a six-fold coincidence of the OR signals from the FEA cards (corresponding to a triple coincidence of both ends of the chambers), whose signals are combined in a VME custom made trigger module. 
The arrival times of the signals are measured using two commercial TDCs (CAEN V1190 - 64 and 128 ch - 100 ps bin).
Synchronization between telescopes is guaranteed by a GPS unit that provides the event time stamp with precision of the order of 40 ns.
Data acquisition, monitoring and control are managed by a LabVIEW based program.

\subsection{\textit{The track reconstruction algorithm}}
\label{recoAlgo}

Data reconstruction is centrally managed at CNAF. Raw data are first processed to calibrate the telescope. Each readout channel has a time offset which slowly drifts in time, mainly in relation with environmental temperature variations. The drift is slow and does not affect a single run, whose approximate duration is typically around half an hour. Every signal induced on a strip generates two time measurements,  $t_{l}$ or $t_{r}$, corresponding to arrival time of the signal to the edges of the chambers conventionally labelled as "right" and "left". 
The calibration is performed for each strip: the $t_{l_{i}} - t_{r_{i}}$ distribution of the i-th strip is used to calculate the mean value $\overline {t_{l_{i}}-t_{r_{i}}}$ for that strip. The mean is then subtracted to measurements $t_{l_{i}} - t_{r_{i}}$, so that the corrected mean value is set to zero. In this way all time differences on each chamber are equalized.
Once the calibration is done, the corrected time values on one strip are paired to form a hit point. 
\\As already pointed out, transversal coordinate is given by the strip number, while the longitudinal coordinate is given by the difference in the arrival time of the signal to the two chamber edges. A time value is also assigned to the hit, computed as the arithmetic mean of $t_l$ and $t_r$ and thus independent from the hit position. 
Cuts are then applied to exclude non physical hits by constraining their longitudinal position: the hit is rejected if $t_l$ and $t_r$ give an $x$ coordinate larger than 79 cm on one of the two sides wrt. the center of the strip (i.e. we require the hit to be inside the active volume). After all hits are reconstructed, clusters are defined by grouping adjacent hits, if present. Hits (tagged 1 and 2) are clusterized if the following requirements are met:
\begin{itemize}
\item hits are on adjacent strips
\item the differences $\Delta t_1-\Delta t_2$, where $\Delta t_{1,2}= t_{l1,2}-t_{r1,2}$, is below 2 ns
\item the hit time difference $t_{h1}-t_{h2}$, where $t_{h1,2}=\frac{t_{l1,2}+t_{r1,2}}{2}$, is below 2 ns
\end{itemize}

\begin{figure}[htb!]
\centerline{
\includegraphics[width=0.8\columnwidth]{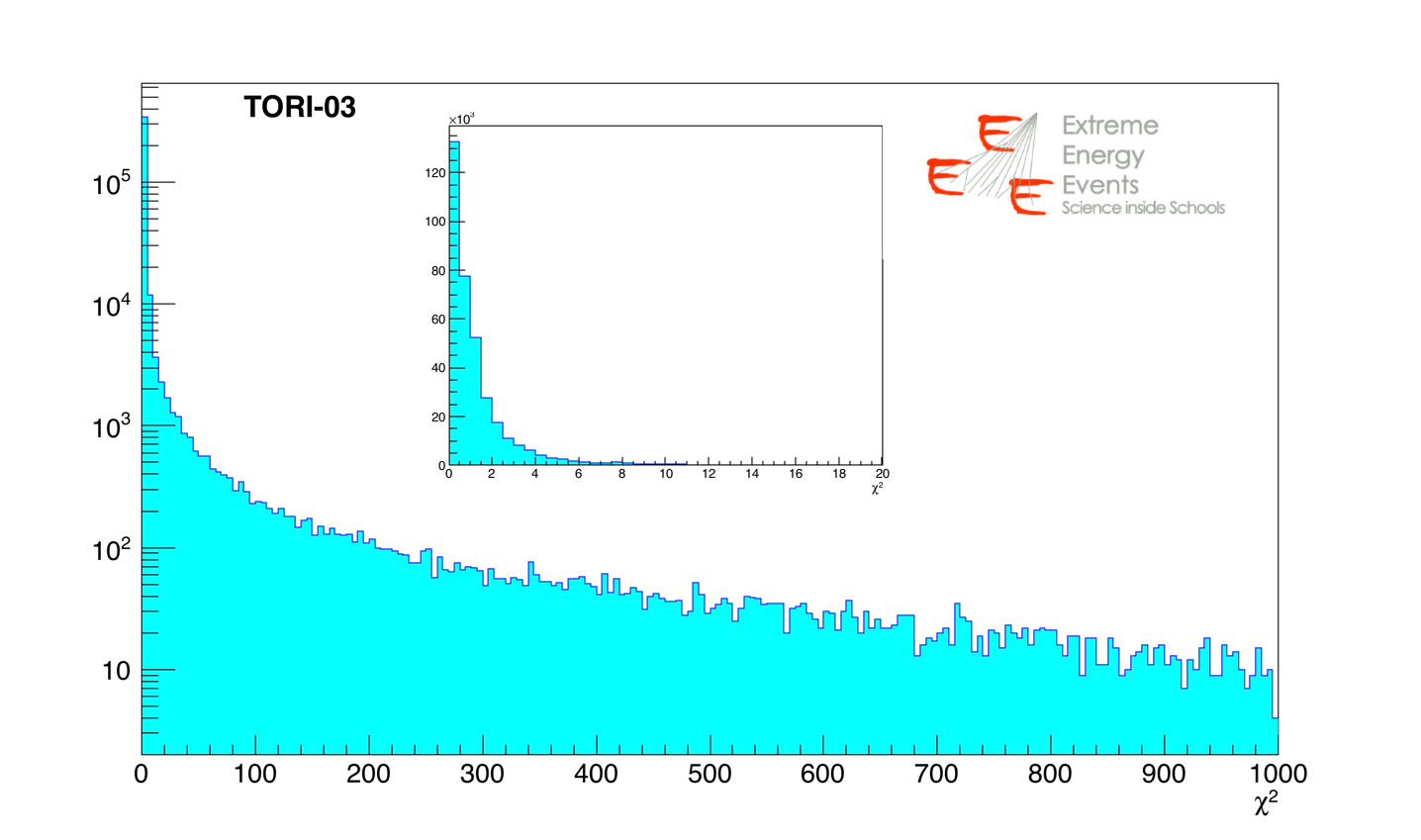}
}
\caption{\footnotesize $\chi^{2}$ distribution for one of the EEE telescopes (TORI-03, located in Piedmont) before any cut is applied. A zoom in the range $\chi^{2}$ = [0-20] is included.}
\label{chi2}
\end{figure}

\begin{figure}[htb!]
\centerline{
\includegraphics[width=0.8\columnwidth]{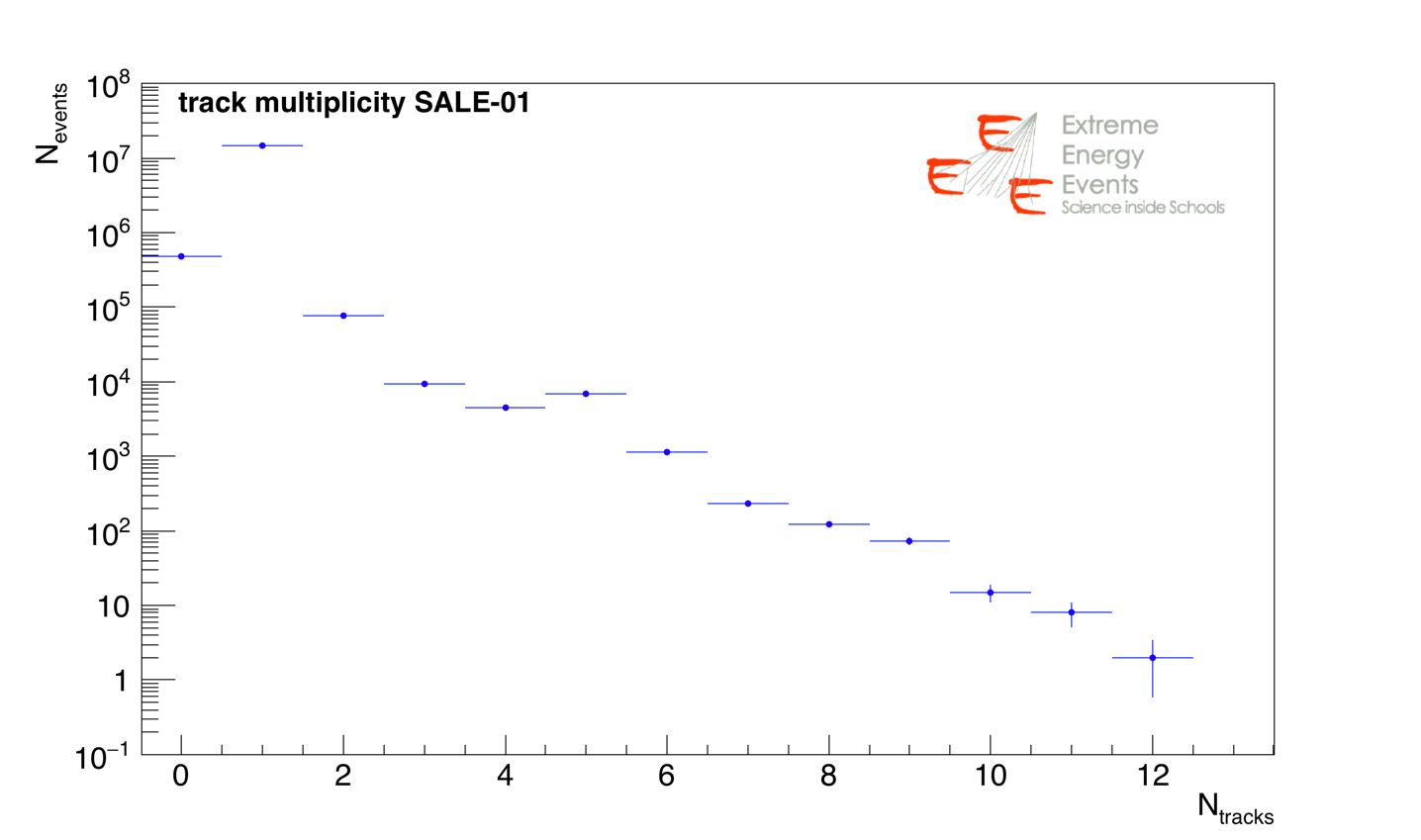}
}
\caption{\footnotesize Track multiplicity distribution of one of the EEE telescopes (SALE-01, located in Campania); the applied cut is $\chi^{2}$ $<$ 5 in this case.}
\label{trkMult}
\end{figure}

The time associated to the cluster is the smallest one.
Finally, track reconstruction is performed. A linear fit of the clusters found in the three chambers is performed and the corresponding $\chi^{2}$ is computed. All possible cluster combinations are used and ordered by their $\chi^{2}$. The track candidates are defined by iteratively selecting the lowest $\chi^{2}$ and removing
the corresponding clusters, continuing up to the point when the  whole set of available clusters has been assigned to a track. At the end a set of tracks with no hits in common is defined and transferred to the output file for analysis.
The track selection for the measurements presented in this paper is done by requiring $\chi^{2}$ $<$ 5 and rejecting events with more than 1 track.
A $\chi^{2}$ distribution before any cut is applied, for 
tracks collected by the EEE station labelled TORI-03, is shown in Fig. \ref{chi2}. The track multiplicity 
distribution for events with a $\chi^{2}$ < 5 is shown in Fig. \ref{trkMult}. 

Some interesting statistics from the four coordinated runs taken so far are listed in table \ref{stats}.
Few considerations are due. The number of tracks wrt. the effective data taking period increased together with the number of telescopes included in the data taking.
The slight decrease of the purity (candidate tracks/triggers) in Run 3 is due to the inclusion of few telescopes not yet optimized in Run 3 and well-performing in Run 4, that started in October 2017.

\begin{table}[htbp]
\centering
\caption{\label{stats}\footnotesize Statistics from the four coordinated runs. The number of active telescopes in Pilot Run, Run 1, Run 2 and Run 3, is respectively 15, 28, 38 and 46. The purity is calculated as candidate tracks/triggers.}
\smallskip
\begin{tabular}{|r|c|c|c|cc|}
\hline
 &Pilot Run&Run 1&Run 2&Run 3&\\
\hline
starting date &$27/10/2014$ & $27/02/2015$ & $07/11/2015$ & $01/11/2016$ &\\
\hline
ending date &$14/11/2014$ & $30/04/2015$ & $20/05/2016$ & $31/05/2017$ &\\
\hline
number of days &$19$ & $63$ & $196$ & $212$ &\\
\hline
tracks/day (M)&$\sim27$ & $\sim53$ & $\sim69$ & $\sim85$ &\\  
\hline
purity (\%)&$75$ & $84$ & $83$ & $80$ &\\  
\hline
\end{tabular}
\end{table}

\noindent The study of the detector performance (described in the next session) has been carried out by selecting a data sample from Run 2 and the recent Run 3. The whole EEE network has been used in the present analysis.

\section{Performance}
\label{performance}

In the next paragraphs the strategy to measure time and spatial resolution will be described. A sample of about 8 billion tracks over 31 billions collected during the Run 2 and Run 3 was used. Time resolution was derived analysing a sample of events collected in Run2 by telecopes TORI-03 (located in Piedmont) and PISA-01 (located in Tuscany).
The main feature of the two telescopes is the fact that they were equipped with a dedicated board which distributed a common clock to the two TDC modules equipping each telescope. This prevented the slow shifts between the two internal TDCs clock, which could spoil the measured time resolution.
In Run 3 the same clock distribution card was installed in all EEE telescopes allowing to asses the time resolution of all telescopes and compare the results with Run 2.
A wider data sample from 46 telescopes has been used to evaluate spatial resolution.
Telescope network time and spatial resolution results of the network are presented in paragraphs \ref{timeres} and \ref{spatres}.
Results of an efficiency measurement, performed without the need of an external detector are shown in paragraph \ref{eff}.
The last paragraph of this section (\ref{Stability}) reports a study on the long term stability of the network, in terms of some quantities like tracking, multiplicity, trigger rate and time of flight.

\subsection{\textit{Time resolution}}
\label{timeres}

The study of the time resolution $\sigma_{t}$ has been performed
by measuring the time information on the upper and lower chambers and using these values to determine the expected time on the middle chamber; this value is then compared with the hit time measured on the middle chamber. The width of the obtained distribution is proportional to the time resolution of the telescope.
Time residuals used for the measurement of the time resolution are therefore defined as:\\

\begin{equation} \label{timeRes}
\Delta t = \frac{t_{top} + t_{bot}}{2} - t_{mid}
\end{equation}

\noindent where $t_{top}$, $t_{mid}$, $t_{bot}$ are the time values for single or clustered hits, as appropriate.

\subsubsection{\textit{Results from RUN 2}}
\label{run2time}

\begin{figure}[htb!]
\centerline{
\includegraphics[width=0.8\columnwidth]{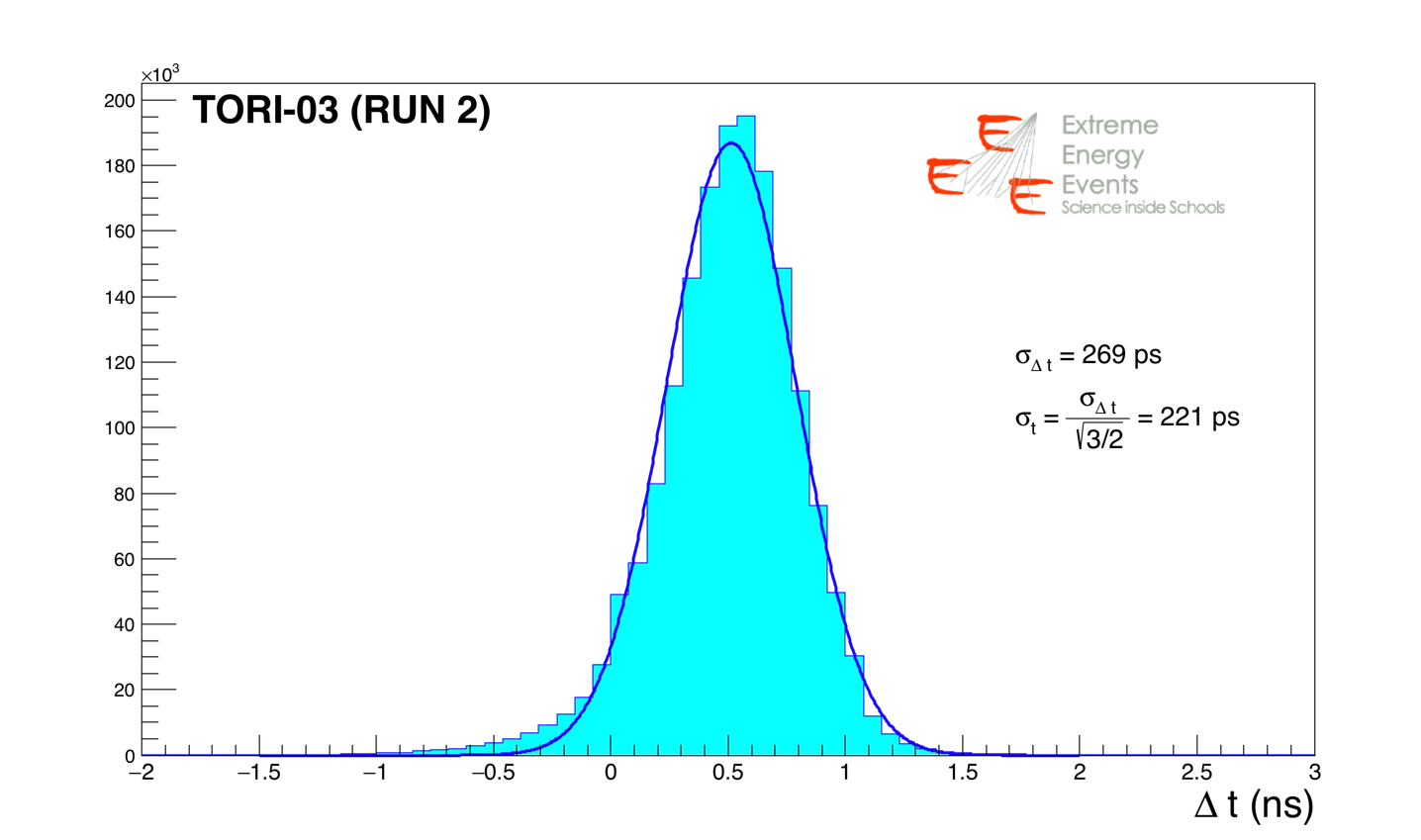}
}
\caption{\footnotesize $\Delta t$ distribution for one of the EEE telescopes (TORI-03), showing the gaussian fit and the time resolution $\sigma_{t}$ = 221 ps. The data used in this case are from Run 2 and the \textit{time slewing} (TS) correction is not applied.}
\label{t_res}
\end{figure}

The $\Delta t$ distribution for TORI-03 (Run 2 data) is shown Fig. \ref{t_res}; the distribution is fitted with a gaussian function whose $\sigma_{\Delta t}$ = 269 ps.  Assuming that the three chambers have similar timing performances, this implies a time resolution $\sigma_{t}$ for the single chamber $\sigma_{t}$ = $\sigma_{\Delta t}/\sqrt{\frac{3}{2}}$ = 221 ps. A similar analysis performed for PISA-01 gave a $\sigma_{t}$ = 270 ps result. It is worth noting that the \textit{time slewing} (TS) correction, explained in details in the next paragraph, is not applied in this case, while is included the analysis performed on Run 3.

\subsubsection{\textit{Results from RUN 3}}
\label{run3time}

\begin{figure}[htb!]
\centerline{
\includegraphics[width=0.8\columnwidth]{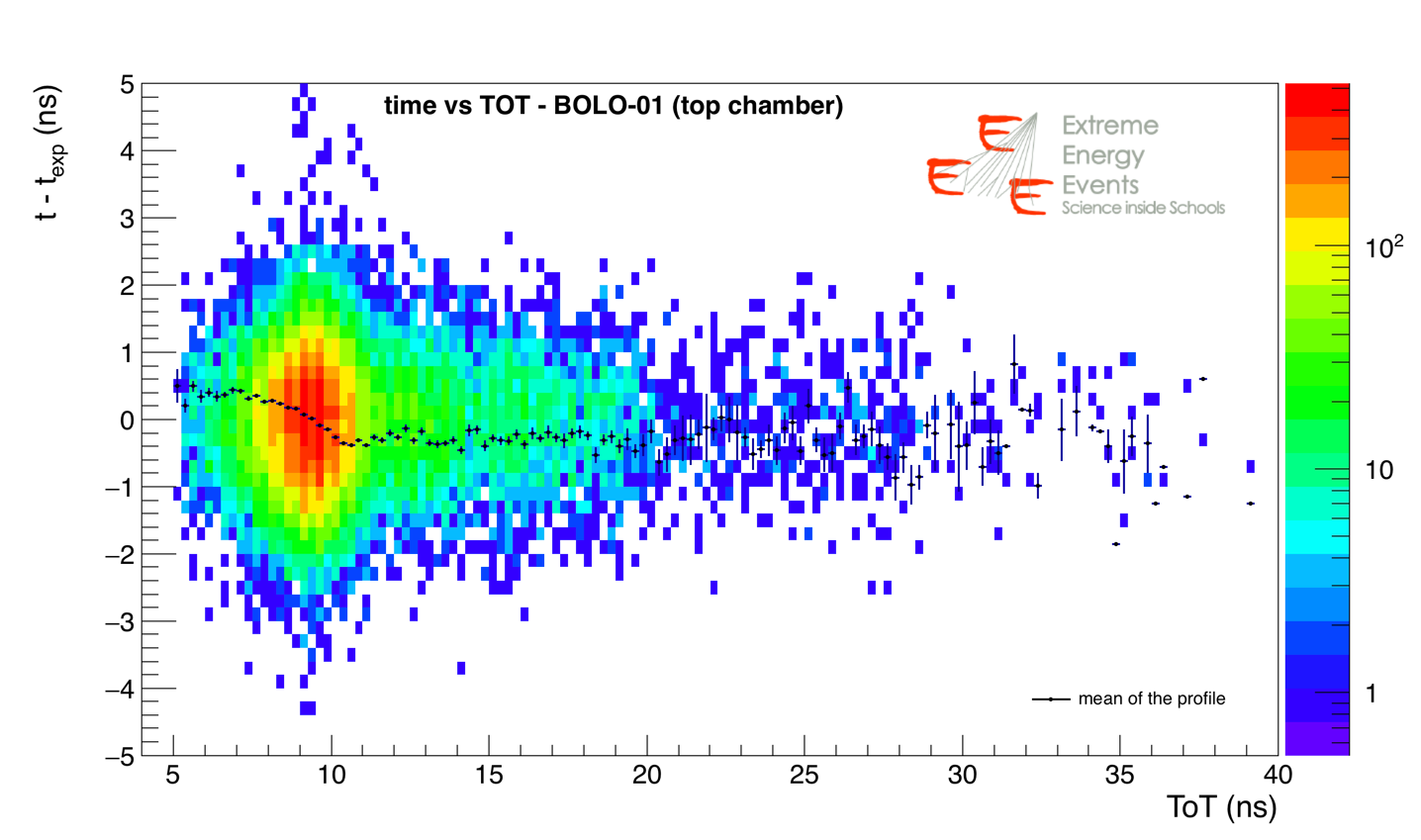}
}
\caption{\footnotesize Correlation between TOT and the difference between the measured time and the time expected on the concerned chamber. The correction to be applied to each hit time is obtained from this distribution (for each single chamber). The distribution of one chamber of the telescope BOLO-01 (located in Emilia Romagna region) is shown here as an example.}
\label{time_slewing}
\end{figure}

The readout electronics used in the EEE project \cite{asic} measures time using so-called \textit{leading edge} discriminators (whose threshold can be chosen in a range [0,1] V and was optimized by setting the value at 500 mV in all FEA) coupled to TDC. The time when the signal becomes lower than the threshold is instead called\textit{trailing edge}.
The Time Over Threshold (TOT) corresponds to the time difference between the times of the trailing and leading
edges and is the time during which the signal remains over the threshold of the signal discriminator.
The hit time depends on the signal amplitude, whose measure is roughly given by the TOT; its jitter can be corrected in order to extract the correct hit time (TS correction).
The correction (performed on each chamber) makes use of the correlation between TOT and $t$-$t_{exp}$, the difference between the measured time $t$ on a specific chamber and the time expected $t_{exp}$ on the same chamber, determined by considering the other two chambers as reference. The procedure for TS correction is standard and can be found in \cite{timeSlewing}.
An example of the mean time vs. TOT distribution for one chamber of one of the telescopes of the EEE network is shown in Fig. \ref{time_slewing}. Each point of this distribution represents the $i-th$ mean value $t_{i,TOT}$ of the $i-th$ bin of the profile histogram; these points  are used to correct the measured time values $t$ ($t_{corr}$ = $t$ - $t_{i,TOT}$). A linear interpolation is performed to get the mean value of $t$-$t_{exp}$ when TOT is between two bins.

\begin{figure}[h!]
\centering
\begin{minipage}[c]{.40\textwidth}
\includegraphics[width=1.28\textwidth]{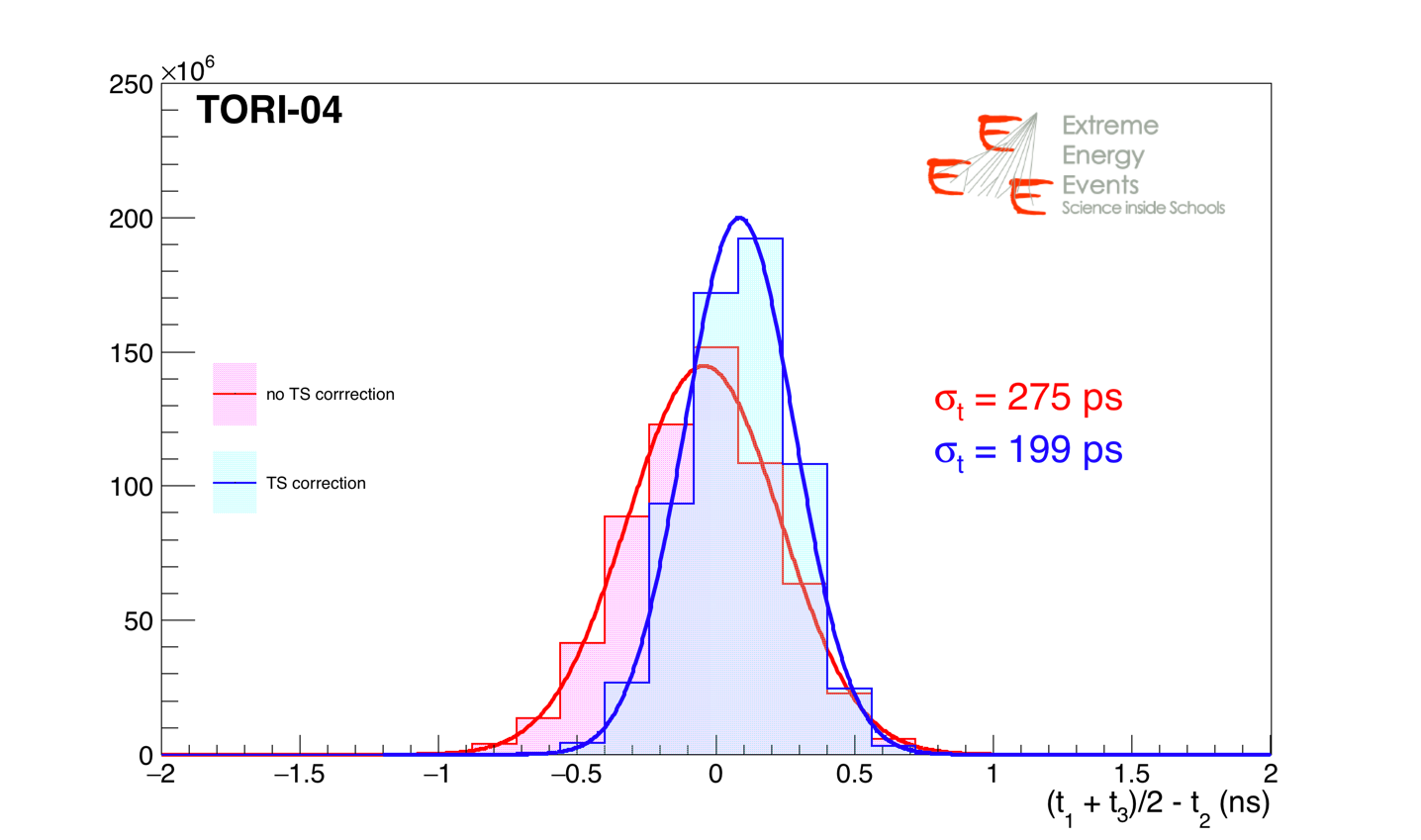}
\end{minipage}%
\hspace{10mm}%
\begin{minipage}[c]{.40\textwidth}
\includegraphics[width=1.28\textwidth]{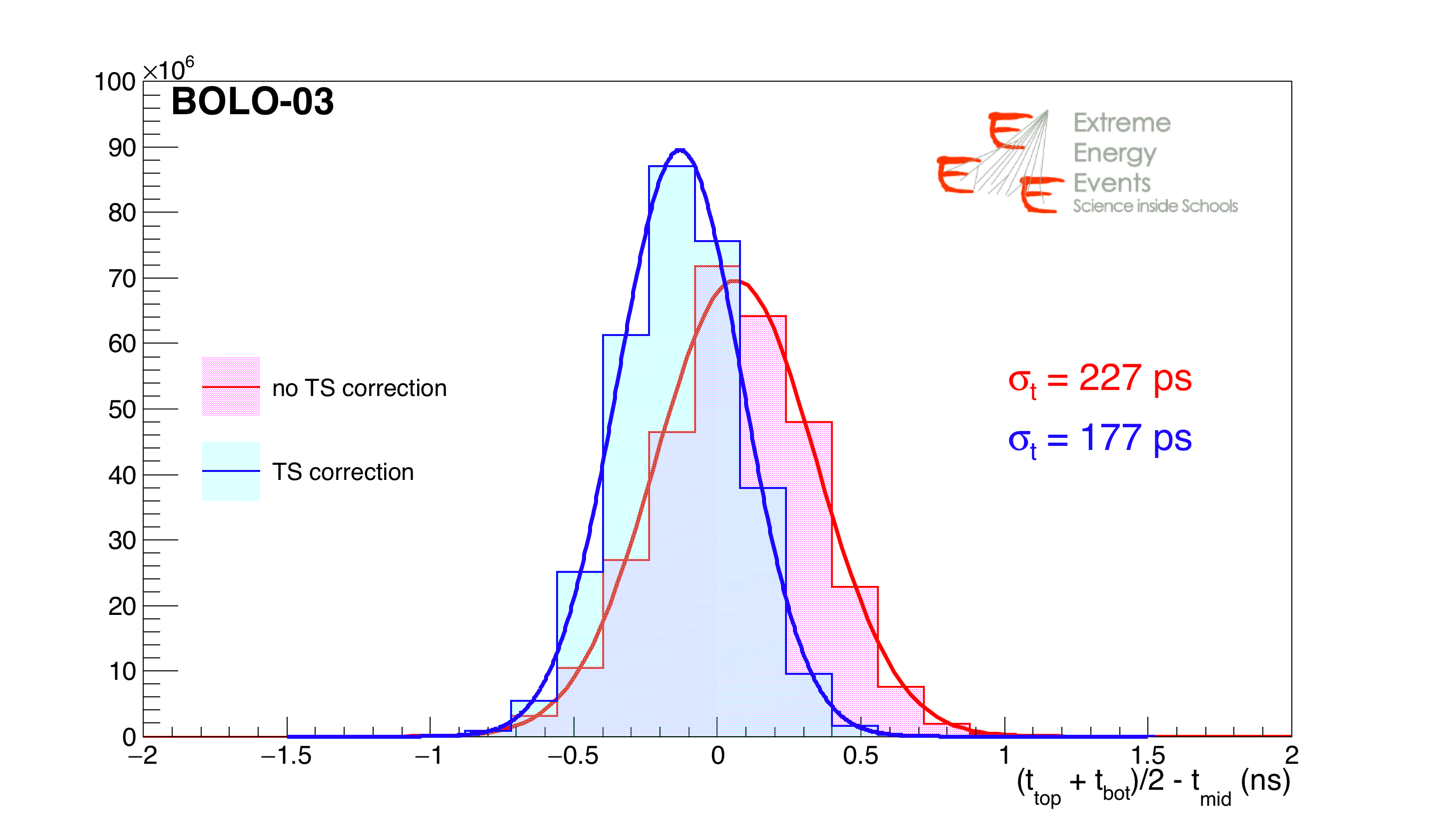}
\end{minipage}
\begin{minipage}[c]{.40\textwidth}
\includegraphics[width=1.28\textwidth]{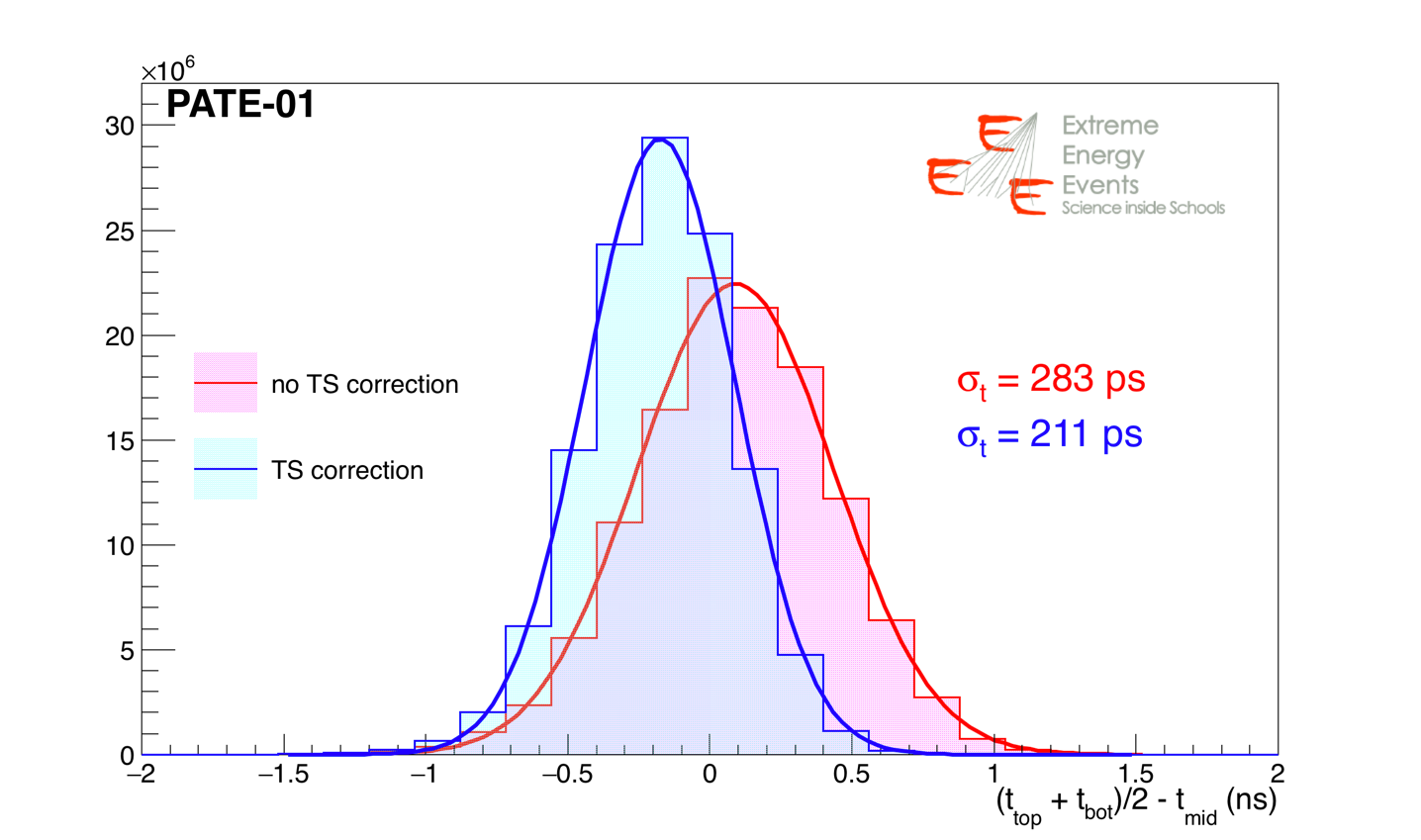}
\end{minipage}%
\hspace{10mm}%
\begin{minipage}[c]{.40\textwidth}
\includegraphics[width=1.28\textwidth]{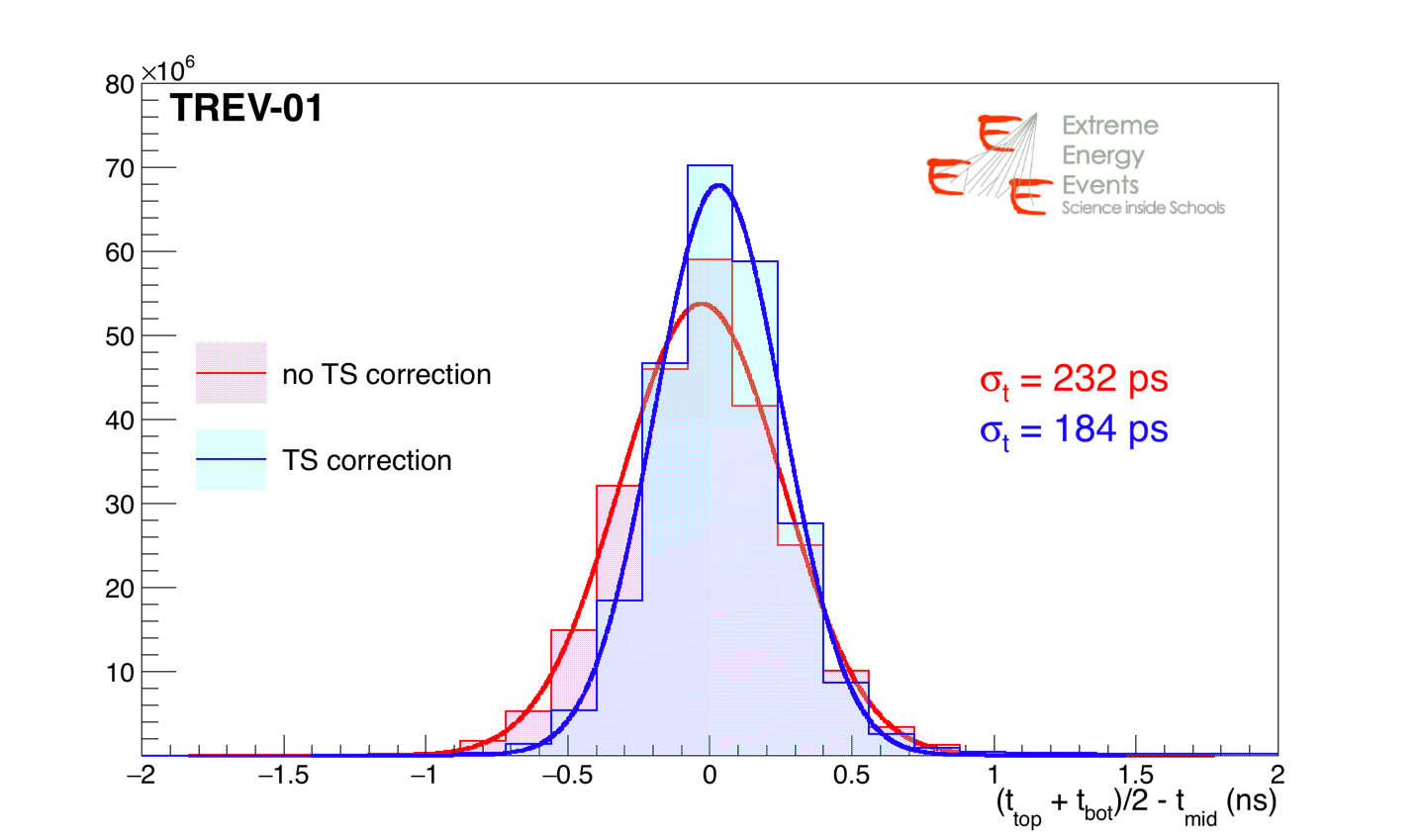}
\end{minipage}
\caption{\footnotesize Time distributions for the telescopes: TORI-04, BOLO-03, PATE-01 and TREV-01 (respectively located in Piedmont, Emilia Romagna, Sicily and Veneto), measured with data taken in Run 3; the distribution and the time resolution before and after TS correction are shown. \label{examples_timeRes}}
\end{figure}

\begin{figure}[htb!]
\centerline{
\includegraphics[width=0.8\columnwidth]
{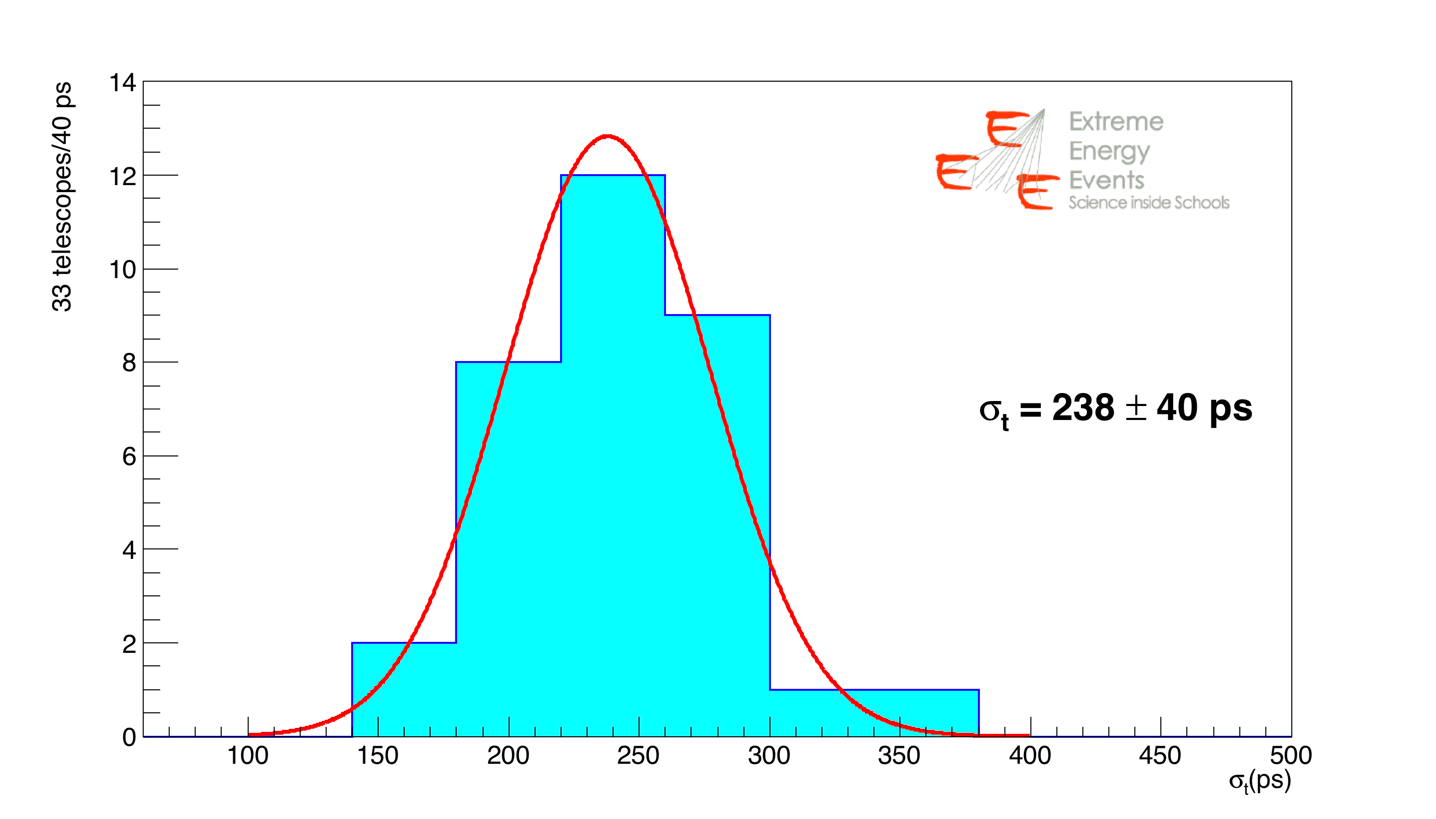}
}
\caption{\footnotesize Time resolution measured with data taken in Run 3, for 33 telescopes; the average time resolution is given by the gaussian fit and is $\sigma_{t}$ = 238 ps.}
\label{timeRes_all}
\end{figure}
Once the correction is determined, it is applied to each hit time and a new time distribution is built by using the corrected time $t_{corr}$, that can be used to measure the time resolution. Some examples are shown in Fig. \ref{examples_timeRes}, where the distributions before and after the TS correction are visible. The measurement has been performed on all the telescopes of the network and a $\approx$$20\%$ improvement is obtained once the correction is applied.
Depending on the telescope the time resolution ranges between 140 and 380 ps. The different values of time resolution for different telescopes can depend on three factors: missing optimization of the detector working point, possible inhomogeneities in MRPC construction, possible not optimal calibration for some strips. The first factor is also related to the dependence of the working point on pressure and temperature; the relative correction is not performed at the moment, but it will be done in Run 4.. 
The other two factors should be less relevant, as the building procedure is robust and tests are periodically performed, as well as the calibration procedure.
A distribution obtained with the values of time resolution from 33 telescopes of the network is shown in Fig.~\ref{timeRes_all}. A gaussian fit gives an average time resolution $\sigma_{t}$ = 238 ps, with a sigma of 40 ps.
This resolution is  within expectations and totally compatible with Run 2 results and with EEE specifications.
It can be compared with the value measured at the beam tests performed in 2006 at CERN \cite{beamTest} of 142 ps without TS correction and $\approx$100 ps with correction and $t_{0}$ subtraction; the smaller value is explained by the fact that at a beam-test conditions are well controlled, with a focused, monochromatic and collinear beam monitored with a set of MultiWire Proportional Chambers (MWPC) and scintillators.
The working points optimization for the next data taking is expected to improve time resolution.

\subsection{\textit{Spatial resolution}}
\label{spatres}
Spatial resolution is obtained by studying the distributions of the particle impact points in the three MRPC.
It has been evaluated by measuring the
spatial information on the upper and lower chambers and by using these values to
determine the expected position on the middle chamber, both in the $xz$ and $yz$ planes ($z$ being the coordinate orthogonal to the chamber plane).
This value is then compared with the hit measured on the middle chamber. The residuals used to measure space resolution are therefore defined as:\\ 

\begin{equation} \label{spaceRes}
\Delta x,y = \frac{x,y_{top} + x,y_{bot}}{2} - x,y_{mid}
\end{equation}
\\ 
Assuming the same space resolution in the three chambers, the space resolution along the strip (longitudinal resolution) of a single chamber can be calculated as
$ \sigma_{x} =  \sigma_{\Delta x}/\sqrt{\frac{3}{2}}$ and along the short side (transverse resolution) as 
$ \sigma_{y} =  \sigma_{\Delta y}/\sqrt{\frac{3}{2}}$.
\subsubsection{\emph{Longitudinal spatial resolution}}
The two signal arrival times, $t_{right}$ and $t_{left}$, are related to the $x_{i}$ coordinate of the hit, to the chamber lenght $L$ 
and to the signal velocity along the strip $v_{drift}$ by the following relation:
\begin{equation} \label{timeHits}
 t_{right}= \Big( \frac{L}{2}-x_{i}\Big)\frac{1}{v_{drift}} ~~~~~~~~~\textnormal{and}~~~~~~~  t_{left} = \Big( \frac{L}{2}+x_{i}\Big) \frac{1}{v_{drift}}
\end{equation}
Therefore the $x$ position is evaluated as an average from both equations, by the difference of the two times:
\begin{equation} \label{xCoordinate}
 x_i = \frac{t_{left}-t_{right}}{2}~v_{drift}  
\end{equation}
where $v_{drift}$ has been assumed to be 15.8 cm/ns.\\
\begin{figure}[htb!]
\centerline{
\includegraphics[width=0.8\columnwidth]{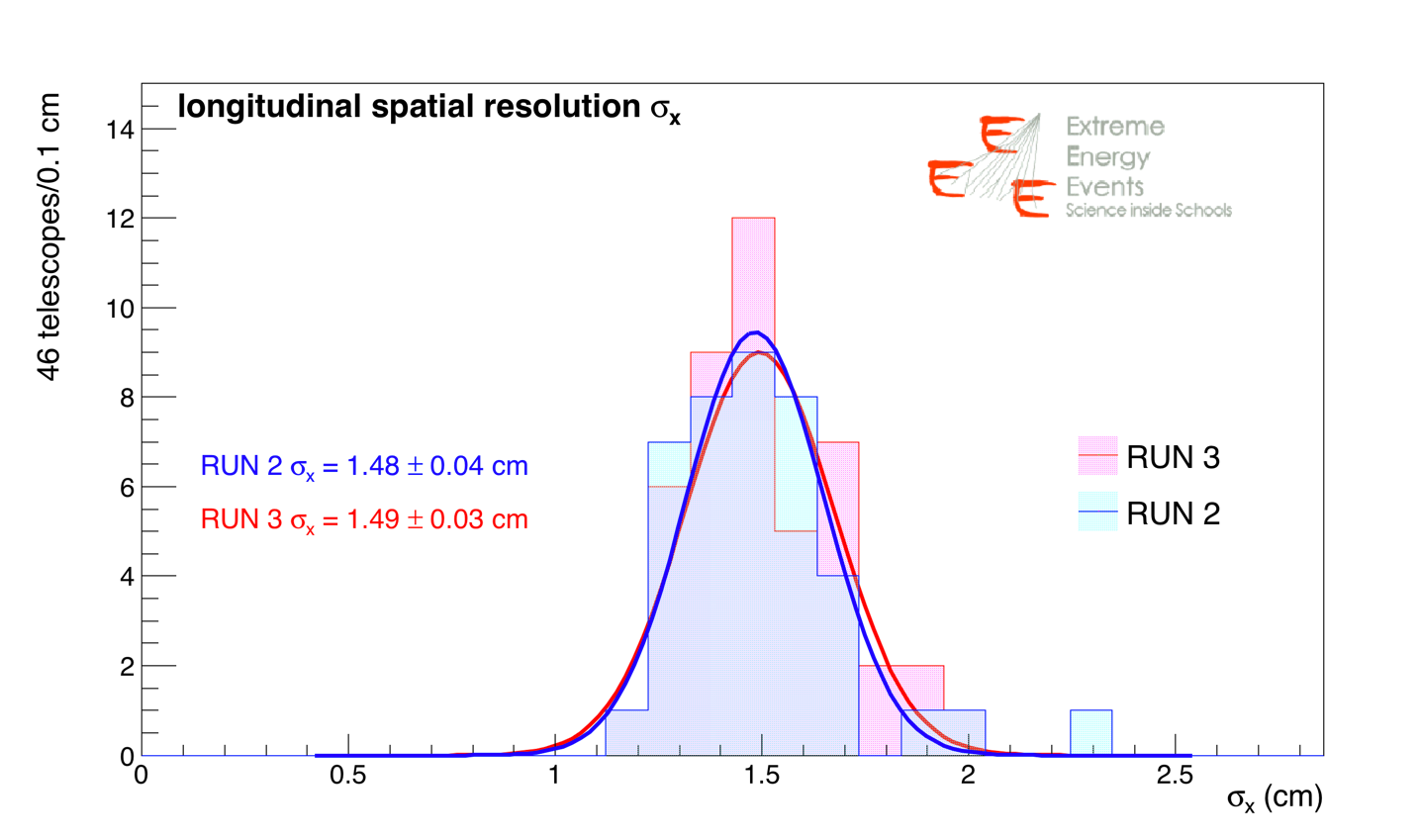}
}
\caption{\footnotesize Longitudinal spatial resolution measured with data taken in Run 2 and Run 3 with 46 telescopes of the EEE network.}
\label{x_res}
\end{figure}
A sample of 2.7 (3.5) billions candidate tracks collected in 30 days from 41 (46)  telescopes in Run 2 (Run 3) has been used for this measurement.
For each telescope the distribution $\mathit{\Delta}$$x$ (see eq. \ref{spaceRes}) has been plotted and used to derive $\sigma_{x}$, with the same strategy already applied to determine the time resolution.
The comparison between the $\sigma_{x}$ distributions from Run 2 and Run 3 are shown in Fig. \ref{x_res}.
 The result from a gaussian fit gives an  average longitudinal resolution of:
  $\sigma_{x_{Run2}} = 1.48 \pm 0.04$ cm and $\sigma_{x_{Run3}}=1.49 \pm 0.03$ cm.
The two results are in agreement, showing the stability of the network across the two runs. 
The longitudinal spatial resolution, measured at the beam test performed in 2006 at CERN previously mentioned, is 0.84 cm; the discrepancy with the measurement reported here is due to a set of missing uncertainty sources and different conditions wrt. beam tests (in a similar way as already explained in Sec. \ref{run3time}).
 \newline
 
\subsubsection{\emph{Transverse spatial resolution}}
The expected transverse spatial resolution is derived considering the pitch of the strips (3.2 cm), $\sigma_{y_{exp}}\sim$ pitch/$\sqrt{12}$ = 0.92 cm.\\
\begin{figure}[htb!]
\centerline{
\includegraphics[width=0.8\columnwidth]{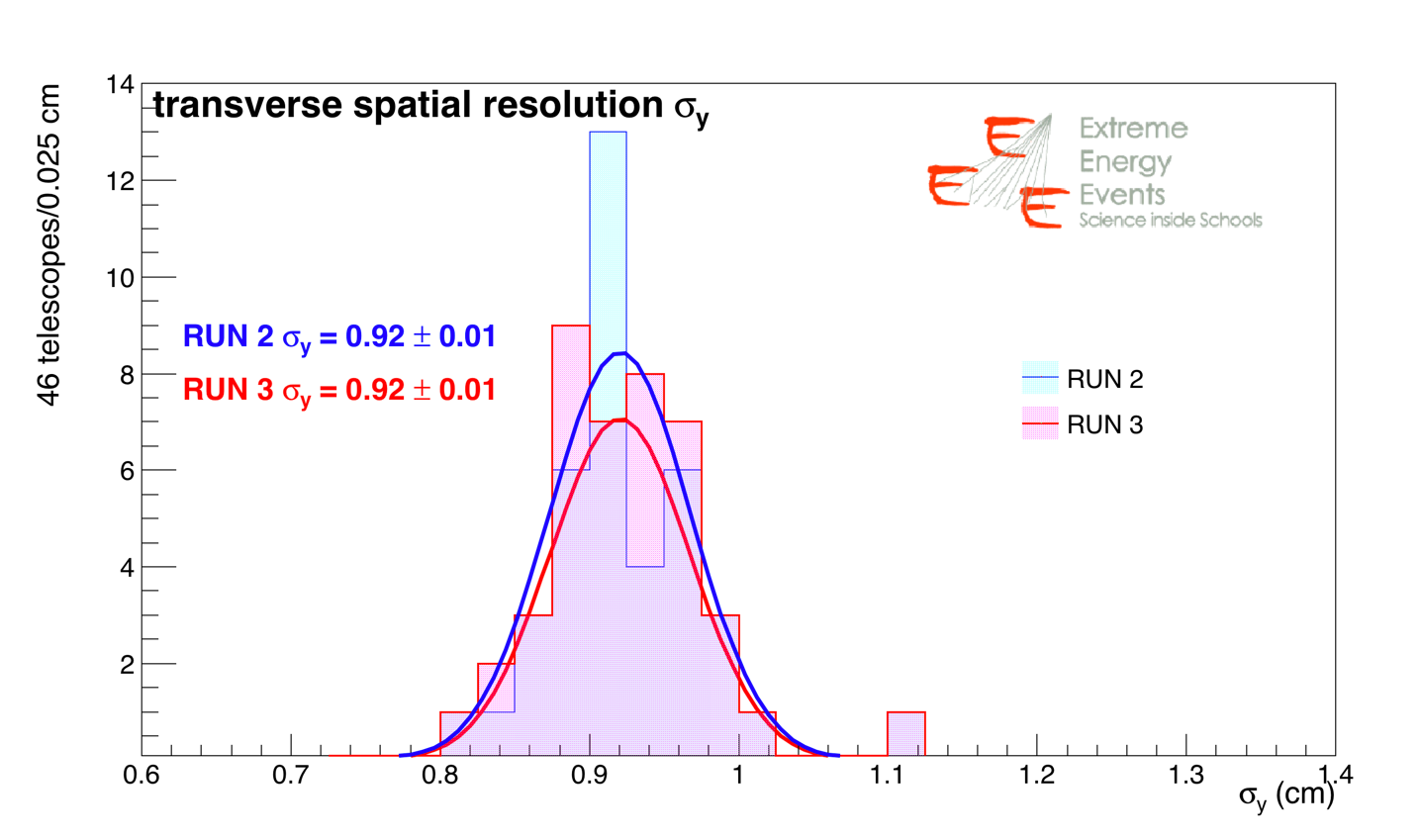}
}
\caption{\footnotesize Transverse spatial resolution measured with data taken in Run 2 and Run 3 for 46 telescopes of the EEE network.}
\label{y_res}
\end{figure}
The spatial resolution in the $y$ direction was measured using the same data samples used to estimate the longitudinal resolution.
For each telescope the distribution $\mathit{\Delta}$$y$ (see eq. \ref{spaceRes}) has been plotted and used to derive $\sigma_{y}$, applying the same strategy used for time resolution evaluation.
The distributions for Run 2 and Run 3 are reported in Fig. \ref{y_res}.
The average resolutions is $ \sigma_{y} = 0.92 \pm 0.01$ cm for both Run 2 and Run 3, in very good agreement with the expectations.

\subsection{\textit{Efficiency}}
\label{eff}

Efficiency curves as a function of the applied voltage have been measured both at CERN, immediately after chamber construction, and after telescopes installation at schools; in most cases these curves have been obtained using scintillator detectors, employed as external trigger, and with additional electronics. \\
Later on, during data acquisition runs, the MRPC efficiency  has been measured without using any additional detector, by using a slightly modified version of the reconstruction code. 
This method allows to check periodically the detectors performance and provides efficiency values, useful for all analysis.\\

\subsubsection{\emph{High Voltage stability}}
\label{HV}
The high voltage (HV) applied to an MRPC, or generally to a RPC-based detector, is a sensitive parameter for all  applications involving absolute particle flux measurements or relative measurements performed over a long time period. The working point of the detector is ideally fixed within the efficiency {\it plateau} region, 300-400 V beyond the knee of the efficiency curve and at the lowest allowed value, in order to limit the chamber spurious counting rate, which usually ranges between 10 and 1000 kHz.
Note that any HV fluctuation beyond 300-400 V (due for instance to changes in temperature and pressure) can move the working point of the chamber in the region left to the knee, where even a few tens of Volt variation corresponds to a significant change in efficiency. An optimal choice of the working point of the EEE telescope is therefore fundamental to allow the EEE telescopes to be sensitive to phenomena involving a few percent particle flux variations, such as solar activity surveys and search for very rare events.\\ 
Indeed since the EEE telescope trigger logic selects events detected by the three MRPC planes, the whole telescope efficiency is $\epsilon_{telescope}$ = $\epsilon_{t}\times\epsilon_{m}\times\epsilon_{b}$, the product of the efficiencies of the top, middle and bottom chambers respectively. Therefore the identification of secondary muons flux variations is strongly affected by the single MRPC efficiency fluctuations.
Assuming the three efficiencies being roughly equal to a common value $\epsilon$, a small single chamber efficiency fluctuation $d\epsilon$ reflects in the telescope efficiency with a fluctuation (at the first order) $\approx \mathrm{3} \epsilon^2 
\,\mathrm{d}\epsilon \approx $ 2.7$\%$, in case of $\epsilon=$0.95 and d$\epsilon\sim$1$\%$.
The typical flux variations connected to a Coronal Mass Ejection on the Sun span from 1-2$\%$ to 6-7$\%$, thus setting the maximum allowed efficiency fluctuation to be well below 1$\%$. \newline
The searches for rare events are even more challenging. The detection efficiency of an EAS by a telescope cluster, composed by $n$ telescopes, is proportional to $\epsilon^{3n}$. The search for long distance correlations between EASs requires the coincidences between 2 clusters (at least 2 telescope each), setting the efficiency for such observations to $\epsilon^{4n}$. By considering $\epsilon=$0.95 one obtains as the overall efficiency for these events $\approx$54$\%$.\\
Temperature and pressure are independent sources of instability for an RPC-based detector, as they affect the mean free path of the charges in the gas volume and thus the detector response \cite{RPC1}.
To mitigate the variation, the EEE collaboration adopted two strategies. The temperature variations are reduced by conditioning the rooms where the telescopes are installed, or an "effective" voltage $HV_{eff}$, as described in \cite{RPC2}, is used. It is defined as :

\begin{equation} \label{effHV}
HV_{eff} = HV \frac{p_{0}}{p} \frac{T}{T_{0}}
\end{equation}
where standard pressure and temperature are set in our case to $p_{0}$ = 1000 mb and 
$T_{0}$=298.15 $K$.

\subsubsection{\emph{Using the outer chambers as a trigger}}
\label{outer}
\begin{figure}[htb]
\centerline{
\includegraphics[width=0.8\columnwidth]{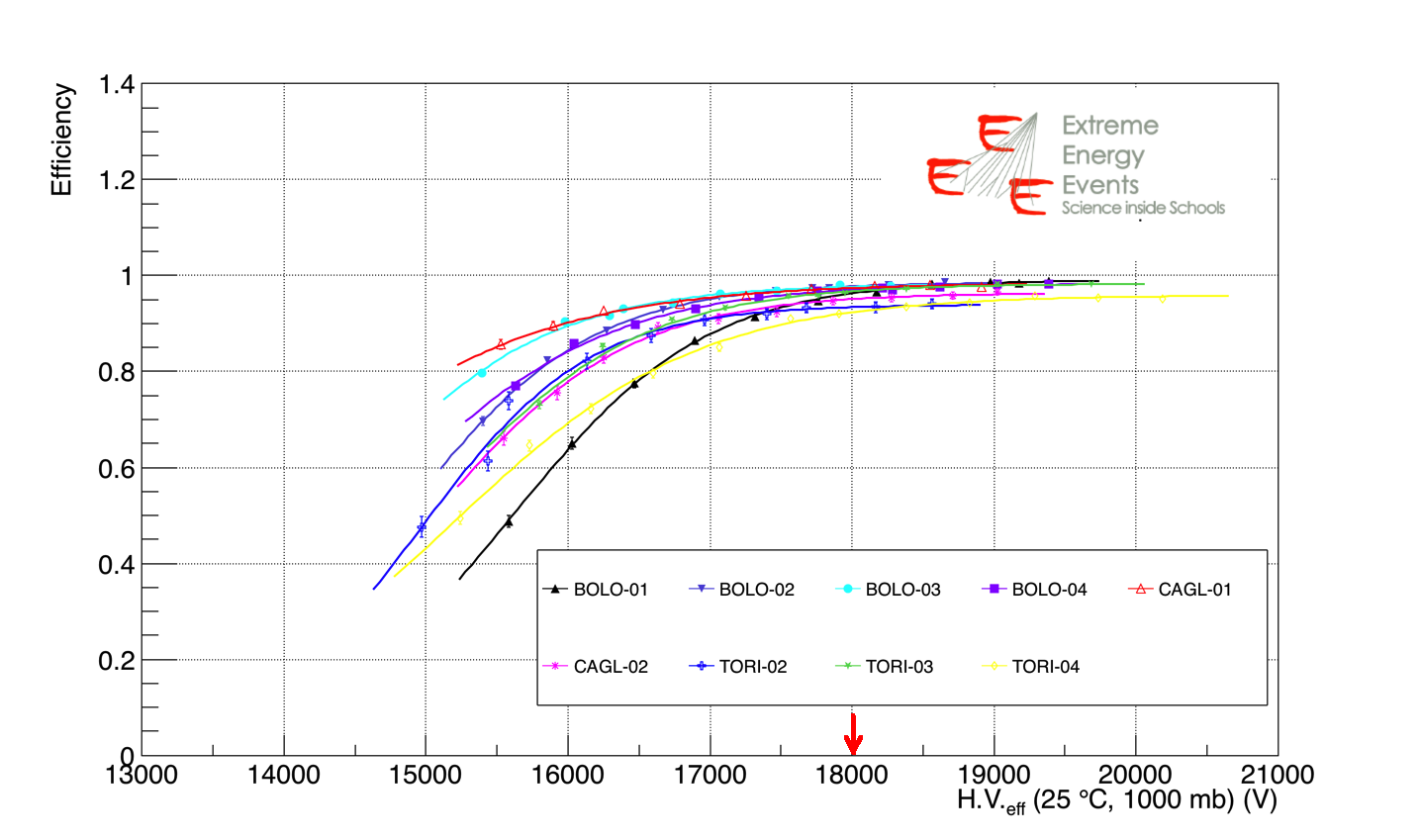}
}
\caption{\footnotesize Efficiency vs. applied HV (corrected for standard $p$ and $T$) of the middle MRPC of 9 EEE telescopes. The red arrow indicates the usual value chosen as working point.}
\label{efficiency}
\end{figure}

The efficiency measurements, whose results are reported here, have been performed by changing from the standard 3-chambers operations to a 2-fold coincidence, excluding the chamber under test from the trigger. 
The two chambers in the trigger are also used for tracking and for selecting events with acceptable values of 1/$\beta$; in particular the 1/$\beta$ distribution is fitted with a gaussian and events are accepted if the 1/$\beta$ value is inside $\pm$ 0.7 from the mean value. Once a track is defined, the procedure requires to check if a hit is present on the chamber under test within a distance of 7 cm wrt. the expected (calculated) position.
An HV scan of the chamber excluded from the trigger is performed, collecting about 150000 events per step. 
An example of the results of these measurements for the middle MPRC of 9 EEE telescopes is shown in Fig. \ref{efficiency}. During the measurements atmospheric pressure $p$ and temperature $T$ were recorded, so efficiency is plotted vs. $HV_{eff}$ (see eq. \ref{effHV} in the previous paragraph).
As shown in Fig. \ref{efficiency}, all MRPC show a similar behavior, with efficiencies reaching almost $100\%$ for an applied voltage larger than 18 kV. 

\begin{figure}[!ht]
\centerline{
\includegraphics[width=0.8\columnwidth]{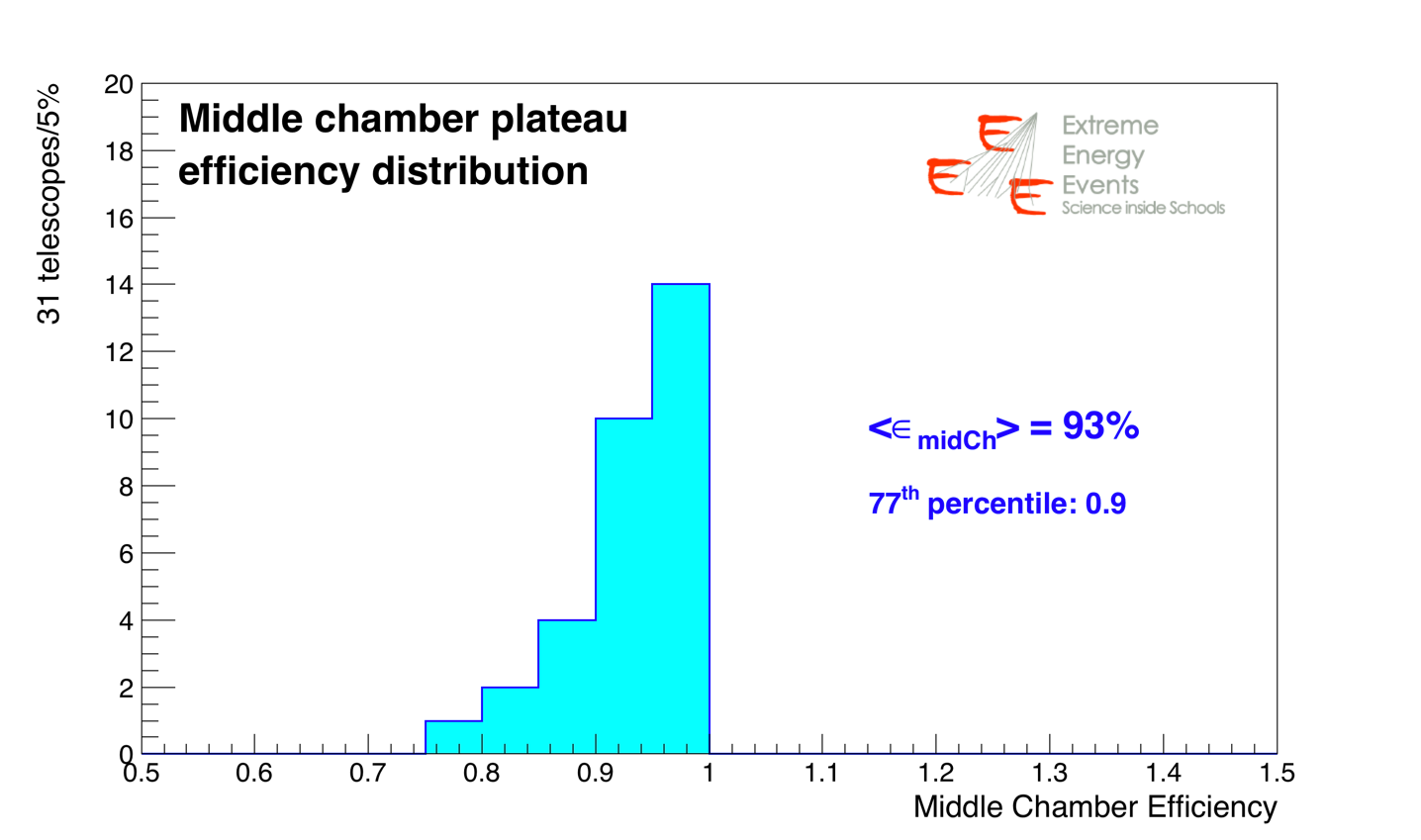}
}
\caption{\footnotesize Distribution of the efficiency obtained at the plateau (corrected for standard $p$ and $T$) of the middle MRPC for 31 EEE telescopes. An efficiency better than 90\% is reached by 77\% of the network.}
\label{global_eff}
\end{figure}
\begin{figure}[htb]
\centerline{
\includegraphics[width=0.8\columnwidth]{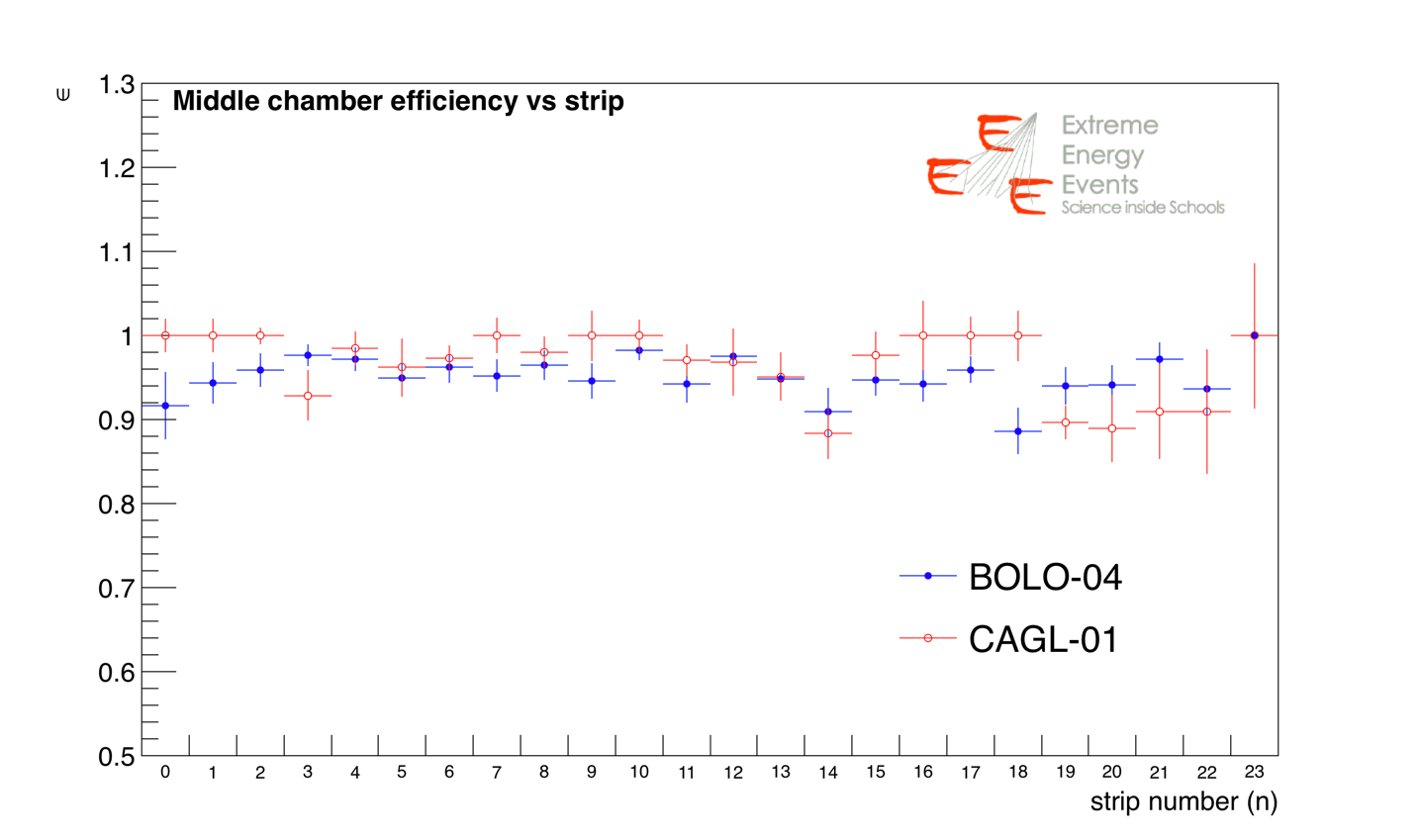}
}
\caption{\footnotesize Efficiency vs. strip for 2 EEE telescopes (located in Emilia Romagna region and Sardinia.}
\label{eff_strip}
\end{figure}

 
This method was applied to the middle chamber of the EEE telescopes, but can be used to measure the efficiency of all the MRPC of a telescope by simply changing the trigger pattern, with the additional care of checking if the predicted hit position lies inside the fiducial area of the chamber under test.
A distribution of efficiency values at the plateau from 31 telescopes (middle chamber) obtained from a three parameters sigmoid function fit to each telescope efficiency curve  is shown in Fig. \ref{global_eff} \cite{RPC2}.
The fit parameters are the efficiency at plateau, the High Voltage at $50\%$ of plateau and the slope of the curve at flush. The average efficiency of the telescope network is around 93\%, compatible with EEE specs and with results of the beam test performed in 2006 at CERN \cite{beamTest}. An efficiency better than 90\% is reached by 77\% of the network, corresponding to 24 telescopes out of 31.
The cause of inefficiency for some telescopes can be related to dead strips and/or MRPC ageing. The efficiency strip by strip for two telescopes of the network, as an example, is shown in Fig. \ref{eff_strip}; the plot shows the efficiency spatial uniformity in the telescopes involved in this measurement. Possible lowering of the efficiency for some strips (taken into account in the reconstruction) are signalled thanks to this detailed measurement.

\subsection{\textit{Long term stability}}
\label{Stability}

Long term performance stability is not easy to achieve with detectors hosted in schools, often far away from the nearest technical support. For telescope monitoring purposes an automatic Data Quality Monitor (DQM) has been created. For each file transferred to CNAF a set of parameters and rates are computed and published online. 
\begin{figure}
	\centering
    \subfloat[$\chi^2$]{
  		\includegraphics[width=0.5\columnwidth]
{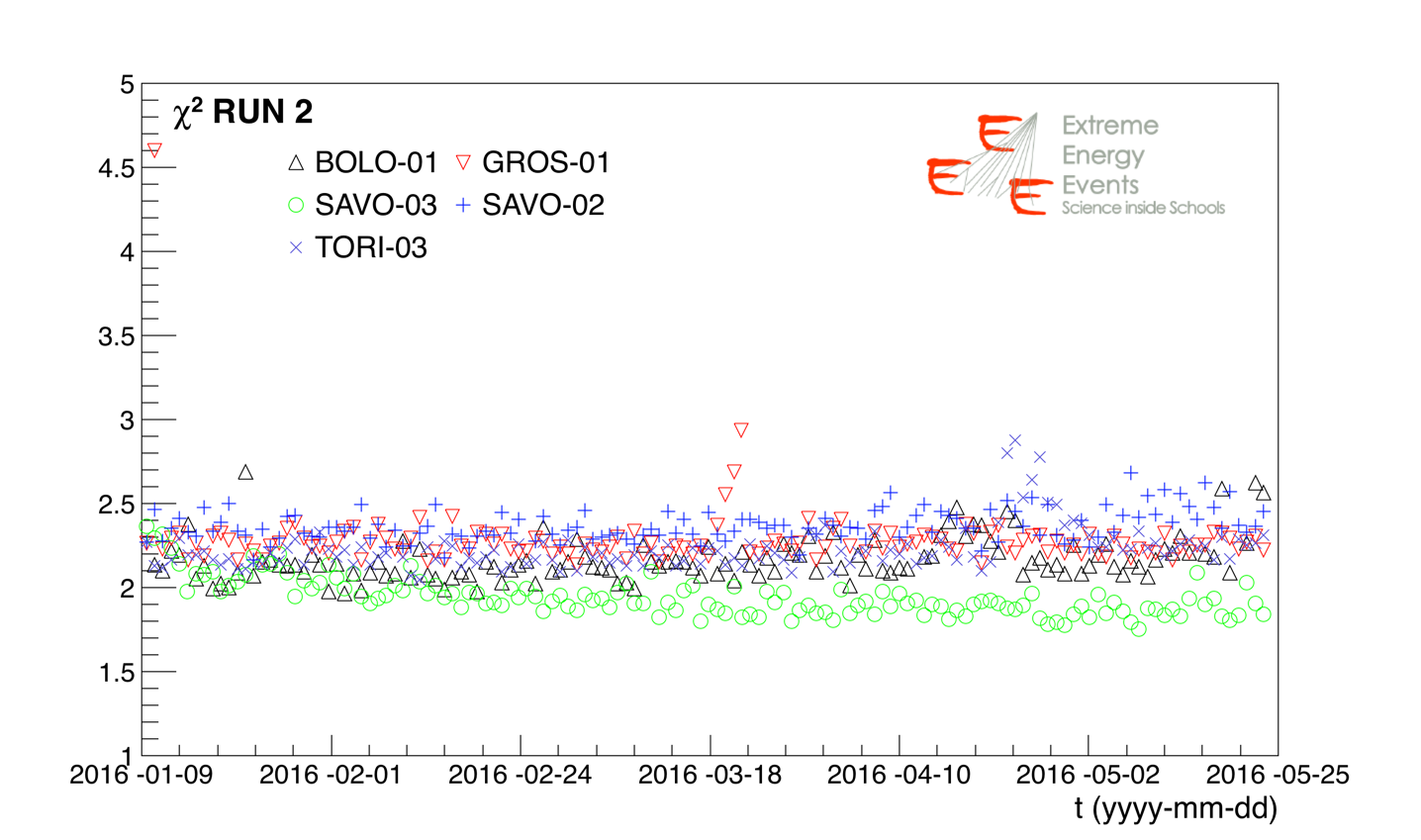}}
	\subfloat[DAQ rate]{
  		\includegraphics[width=0.5\columnwidth]
        {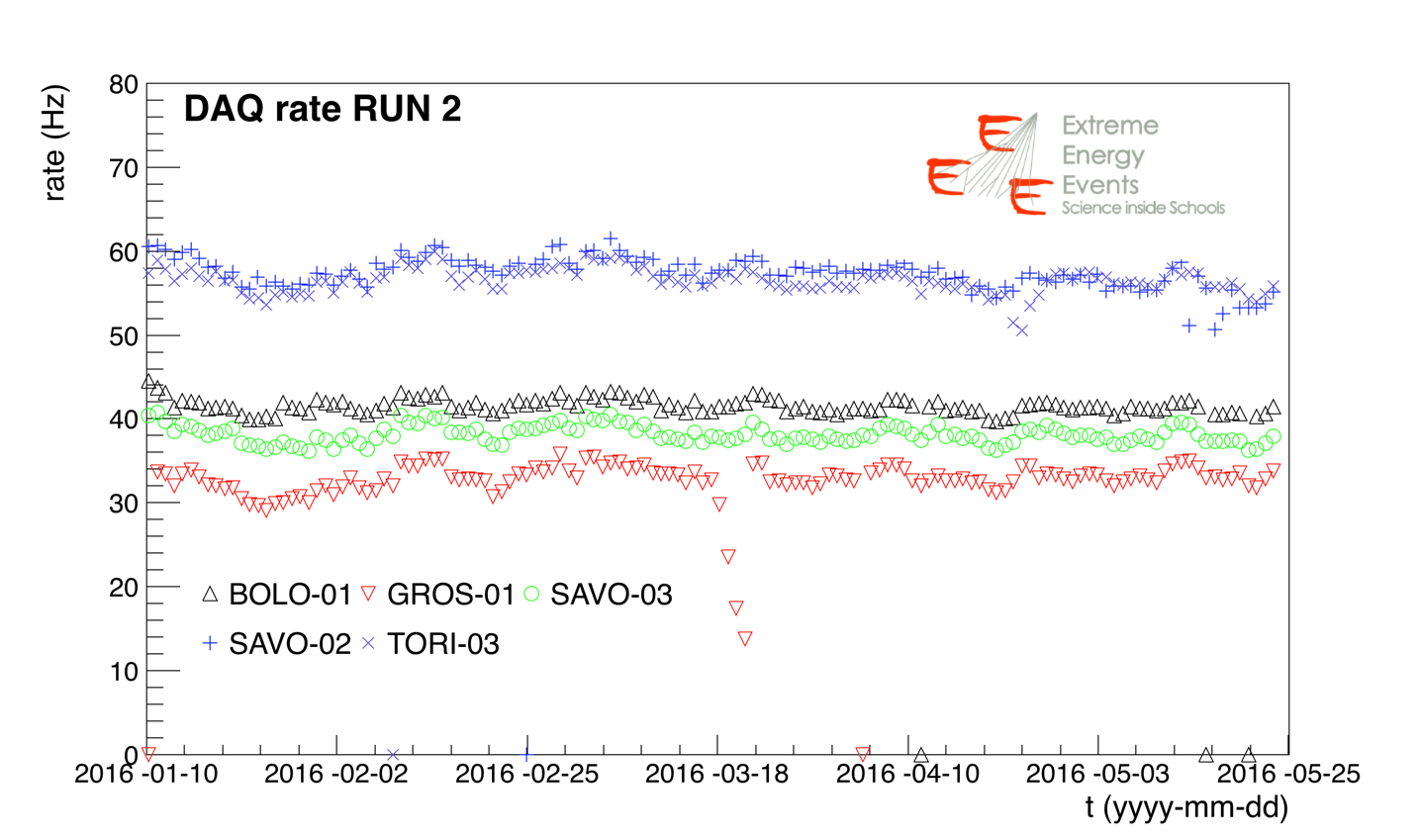}}
	\hspace{0mm}
	\subfloat[Multiplicity]{
  		\includegraphics[width=0.5\columnwidth]
        {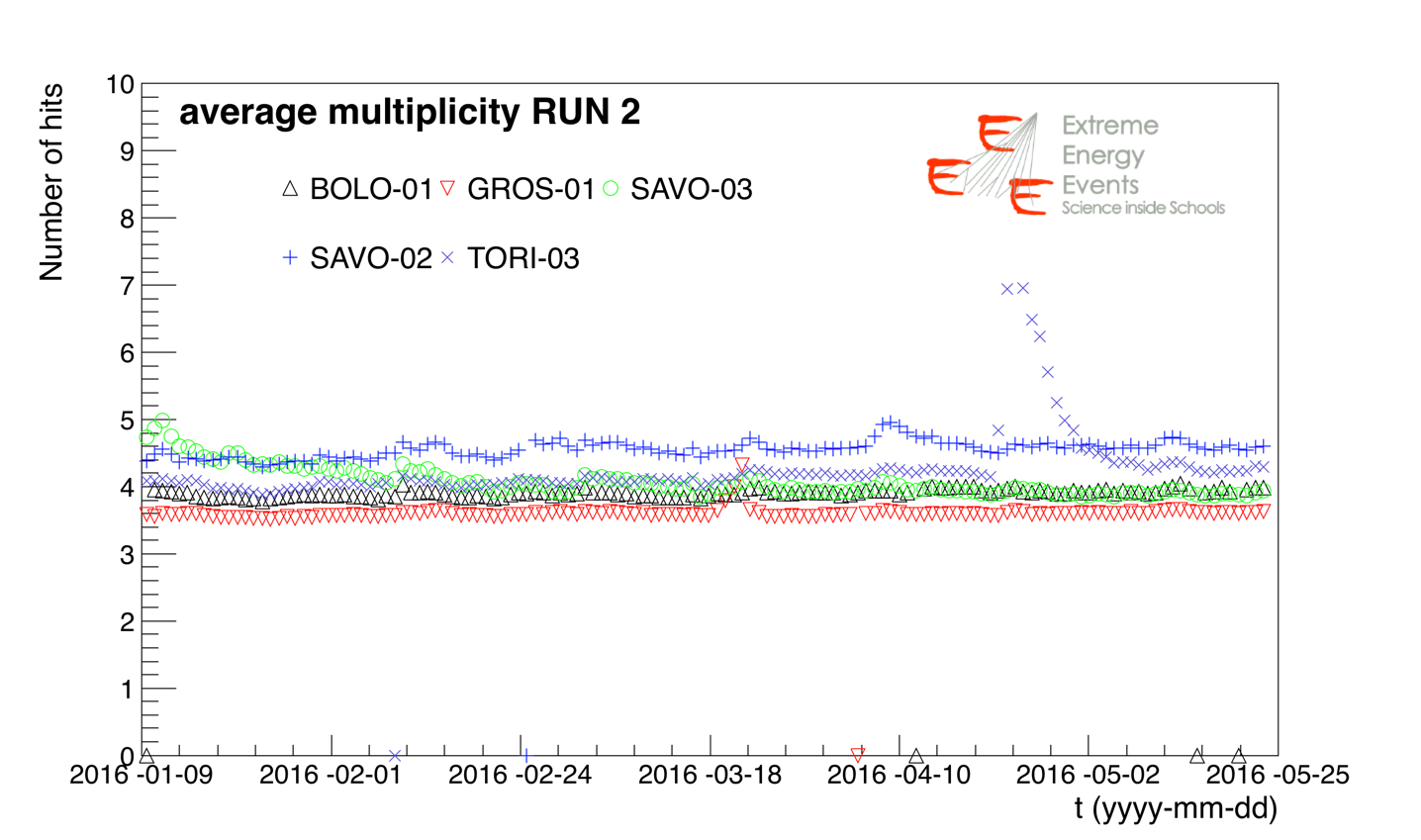}}
	\subfloat[Reconstruction efficiency]{
  		\includegraphics[width=0.5\columnwidth]
        {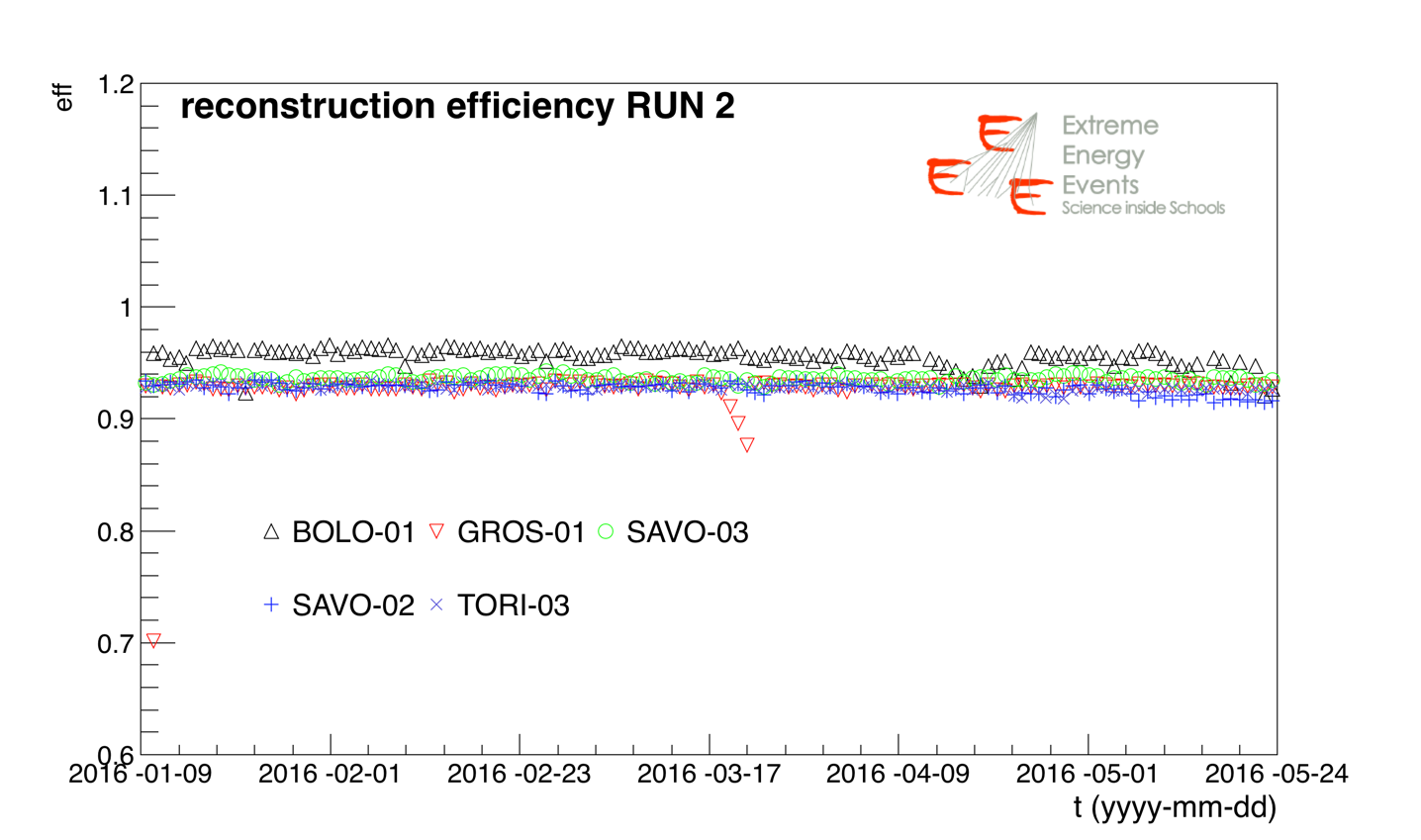}}   
	\hspace{0mm}
	\subfloat[Time of Flight]{
  		\includegraphics[width=0.5\columnwidth]
        {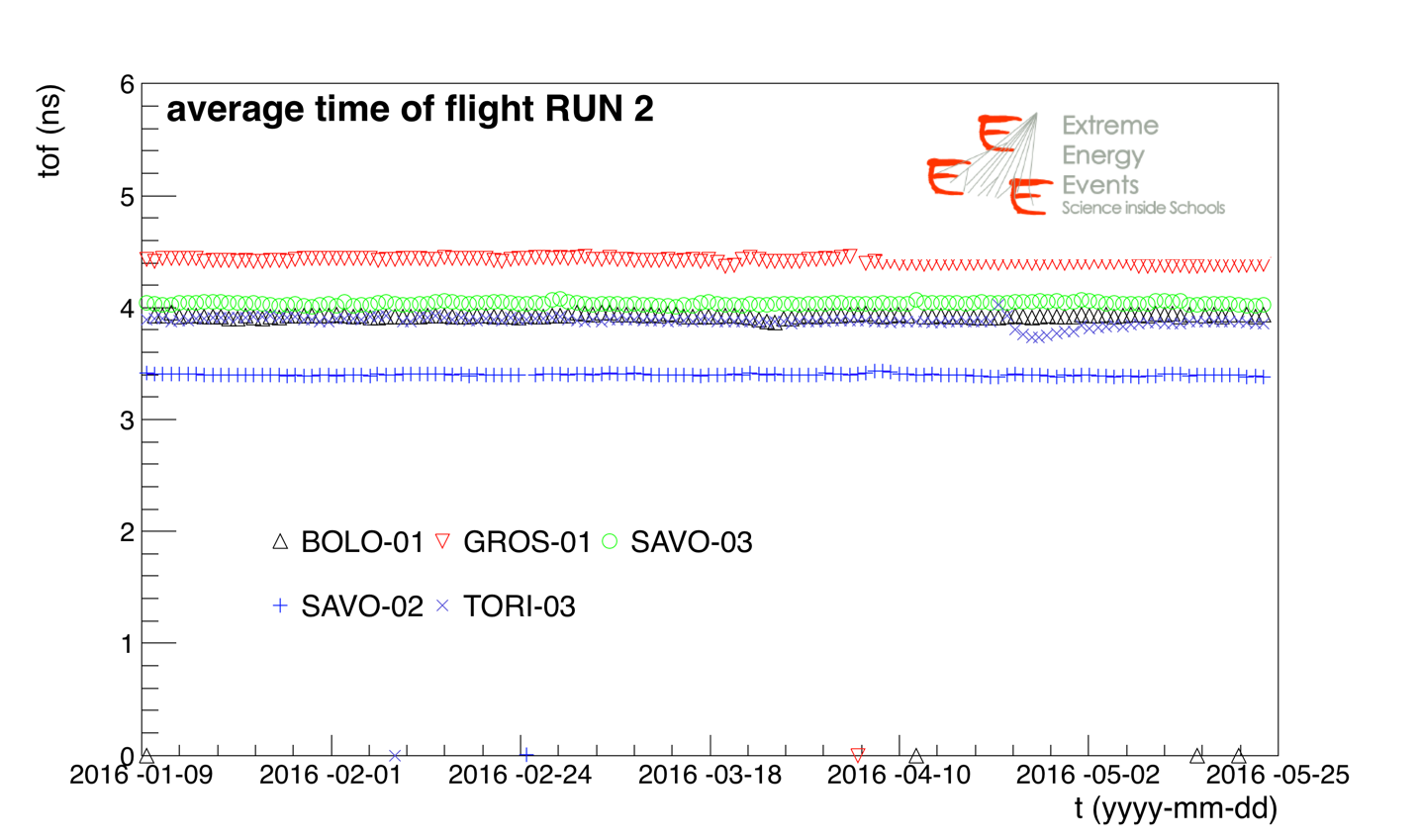}}
	\subfloat[Track rate]{
  		\includegraphics[width=0.5\columnwidth]
       {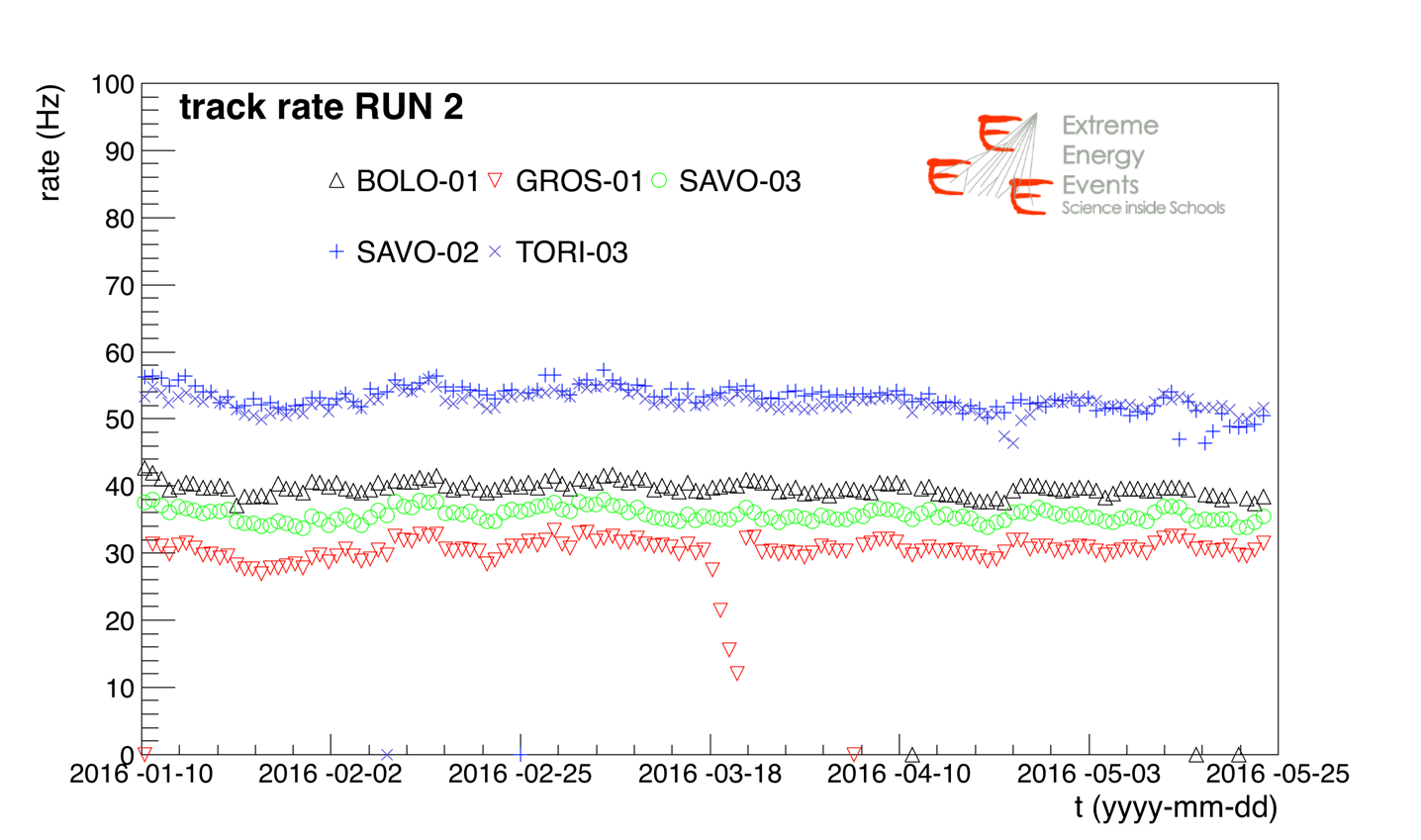}}
	\caption{\footnotesize Run 2 trending plot for 5 EEE stations. Points at zero represent days when the telescope was not operational.}
	\label{Run2dataset}
\end{figure}
Daily report are also automatically generated, illustrating the evolution of the parameters over the last 48 hours. DQM allows for a fast reaction in case one station deviates from the standard behavior. Using the DQM and full reconstruction outputs, it is also possible to extend such trending plots to longer periods. As an example, trends for a selection of relevant quantities are reported in Fig. \ref{Run2dataset} for some telescopes of the EEE network:
\begin{enumerate}
\item $\chi^2$: average tracks $\chi^2$;
\item \emph{DAQ rate}: raw acquisition rate;
\item \emph{Multiplicity}: average number of hits on the three chambers for each event;
\item \emph{Reconstruction efficiency}: percentage of raw events where at least one track candidate has been found;
\item \emph{Time of Flight}: average tracks TOF between top and bottom chambers;
\item \emph{Track rate}: rate of events with at least one candidate track.

\end{enumerate}

\begin{figure}
	\centering
    \subfloat[]{
  		\includegraphics[width=0.5\columnwidth]{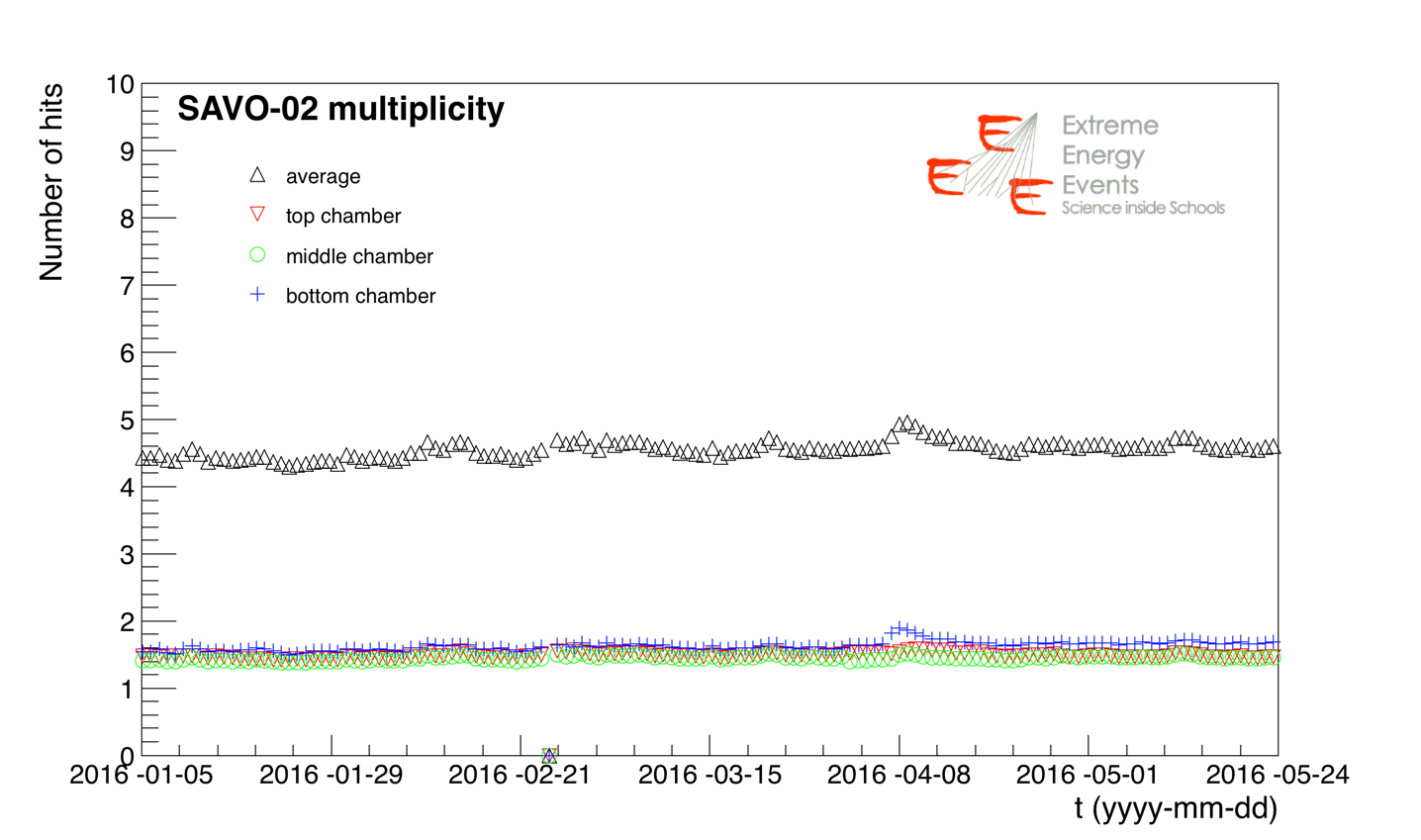}}
	\subfloat[]{
  		\includegraphics[width=0.5\columnwidth]{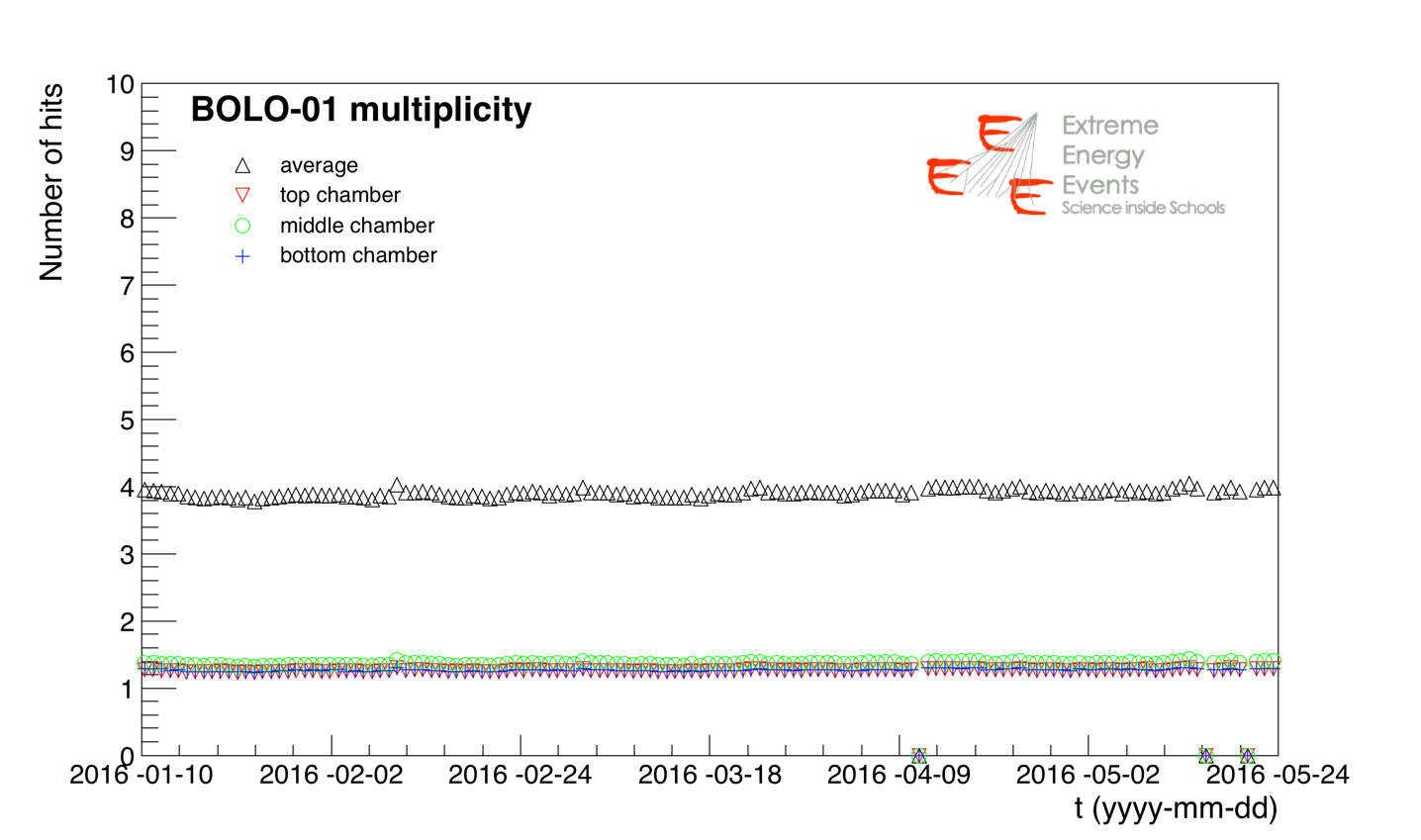}}
      \hspace{0mm}
	\subfloat[]{
  		\includegraphics[width=0.5\columnwidth]{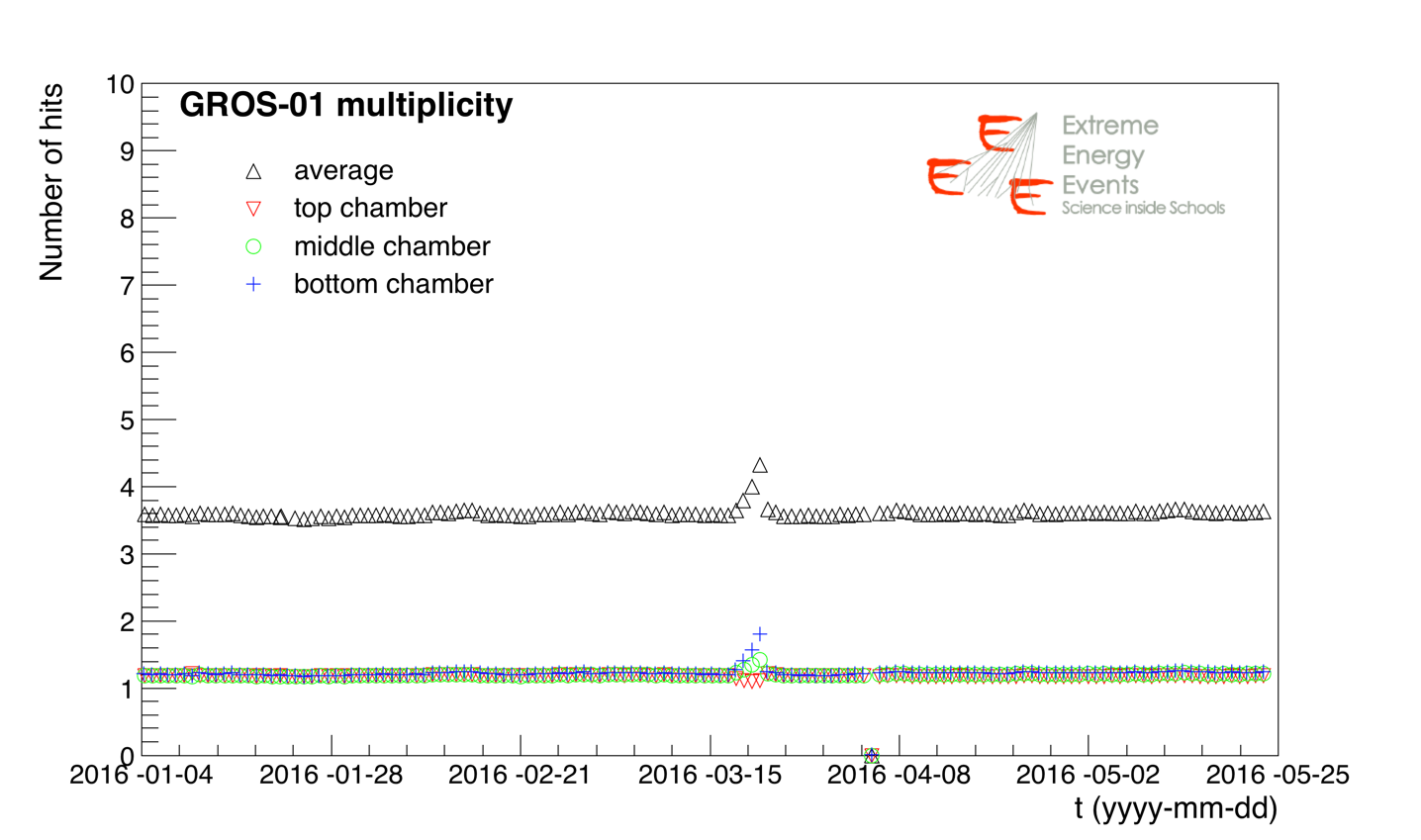}}
	\subfloat[]{
  		\includegraphics[width=0.5\columnwidth]{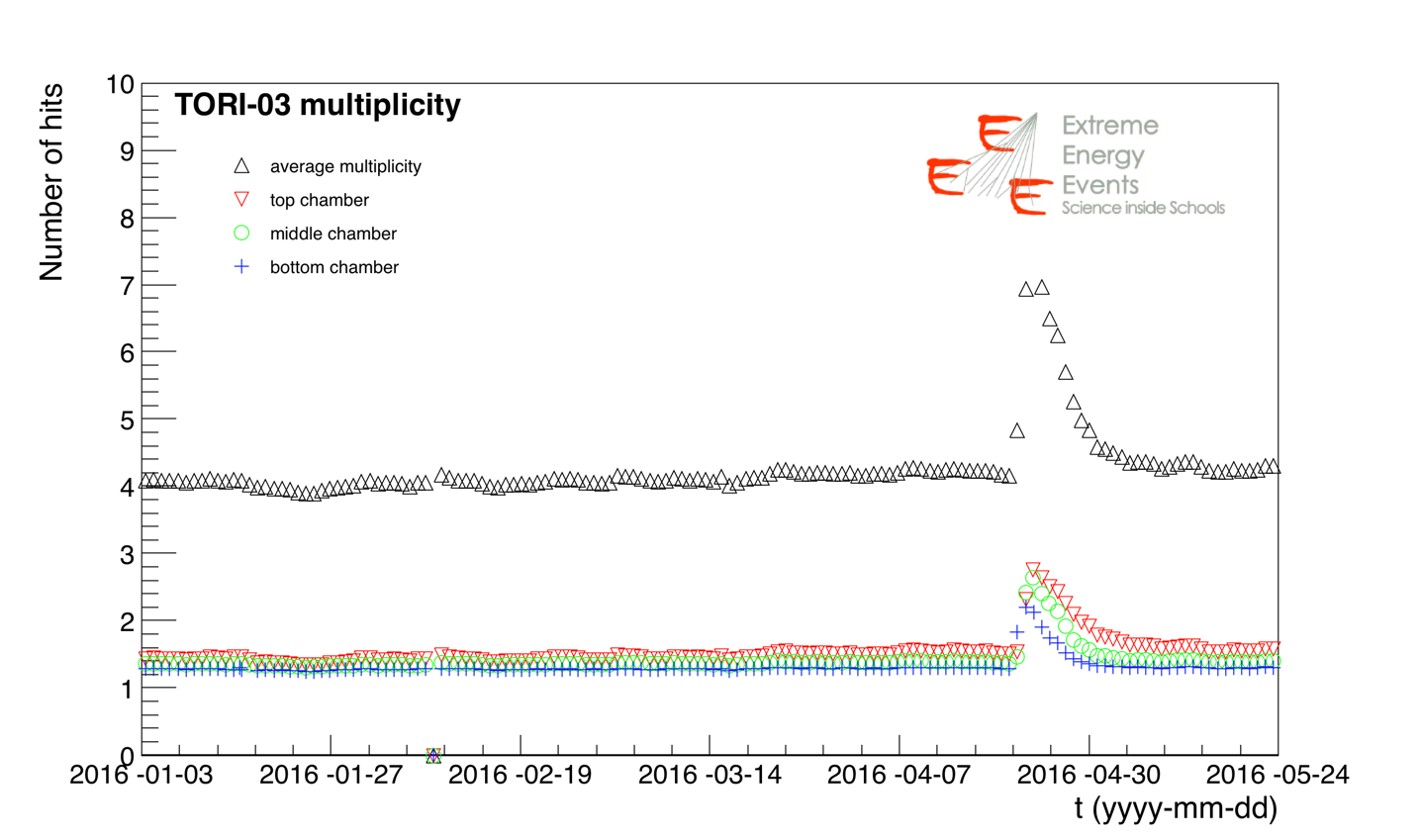}}
	\caption{\footnotesize Average and per chamber hit multiplicity of SAVO-02, BOLO-01, GROS-01 and TORI-03 (located respectively in Liguria region, Emilia Romagna region, Tuscany and Piedmont); a sample from Run 2 data has been used.}
	\label{Run2Mult}
\end{figure}
\begin{figure}
	\centering
    \subfloat[$\chi^2$]{
  		\includegraphics[width=0.5\columnwidth]
        {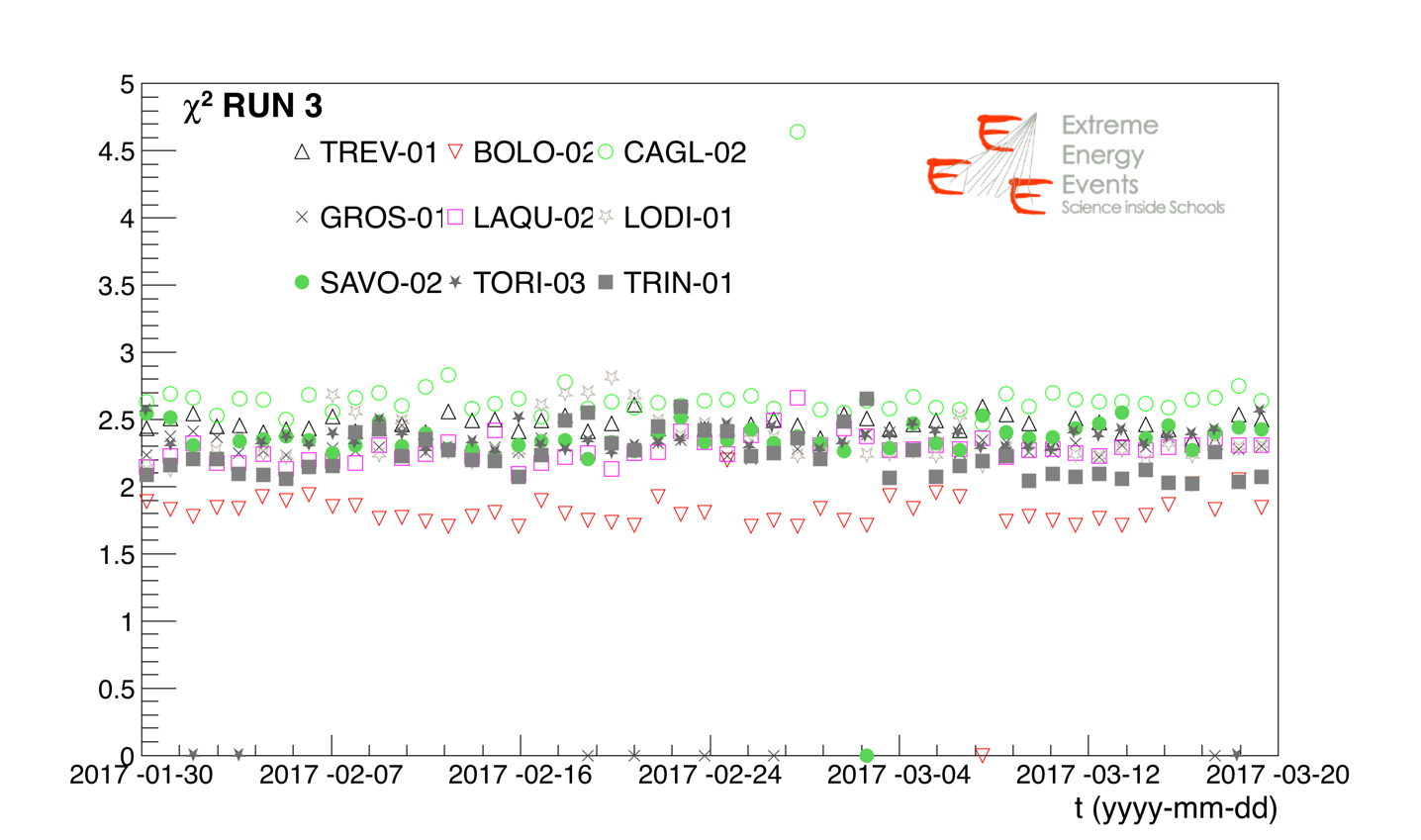}}
	\subfloat[DAQ rate]{
  		\includegraphics[width=0.5\columnwidth]
        {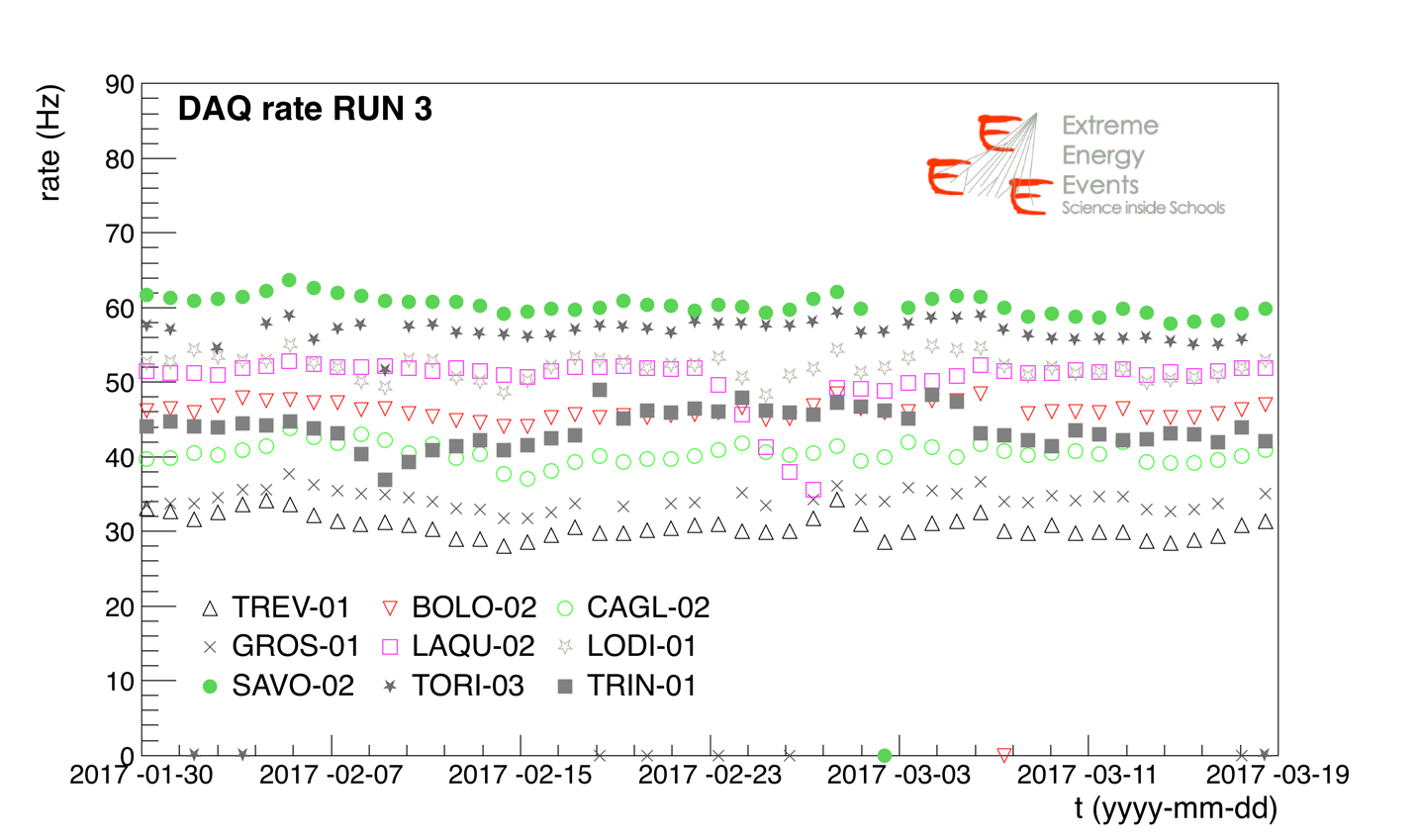}}    
	\hspace{0mm}
	\subfloat[Multiplicity]{
  		\includegraphics[width=0.5\columnwidth]
       {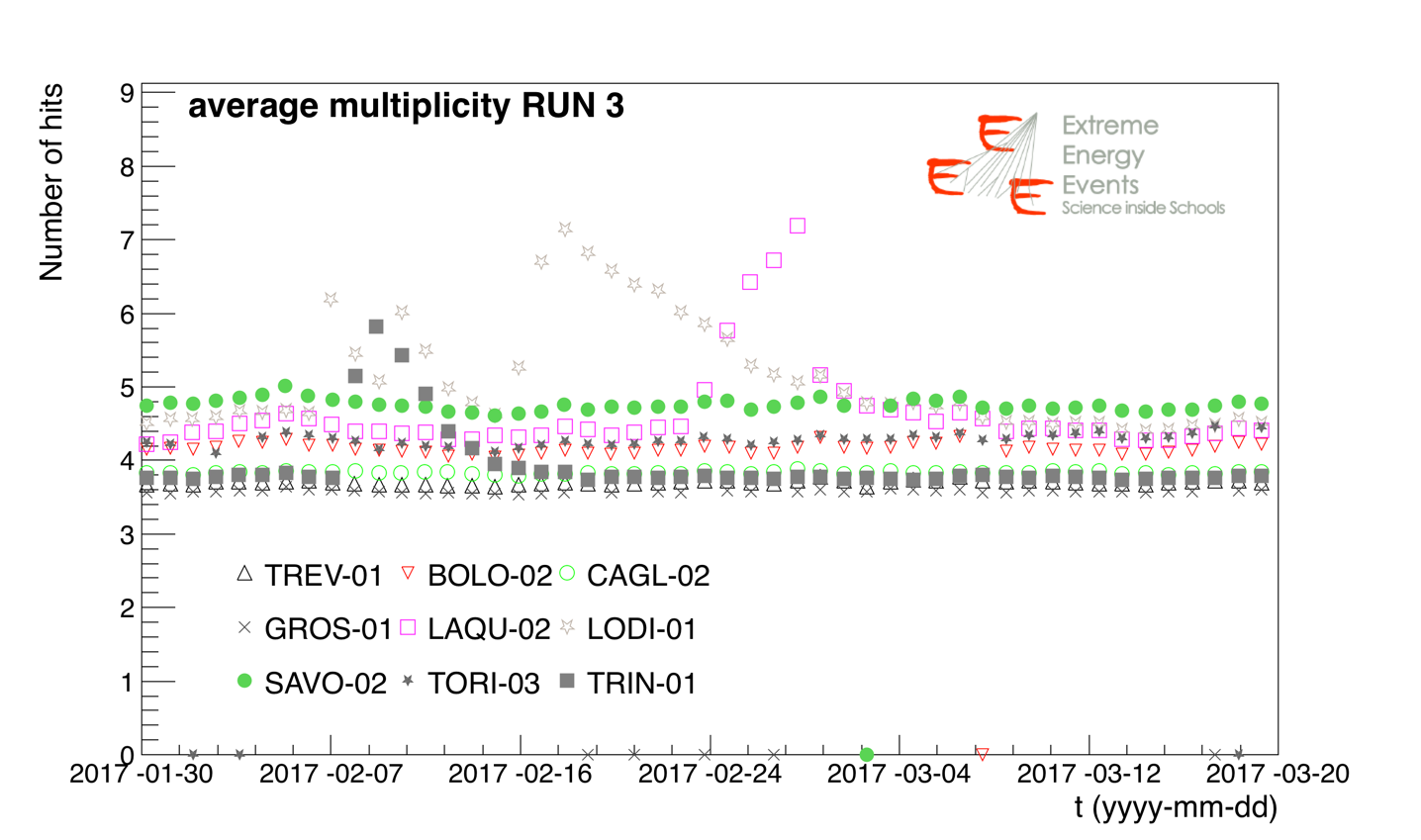}}
	\subfloat[Reconstruction efficiency]{
  		\includegraphics[width=0.5\columnwidth]
       {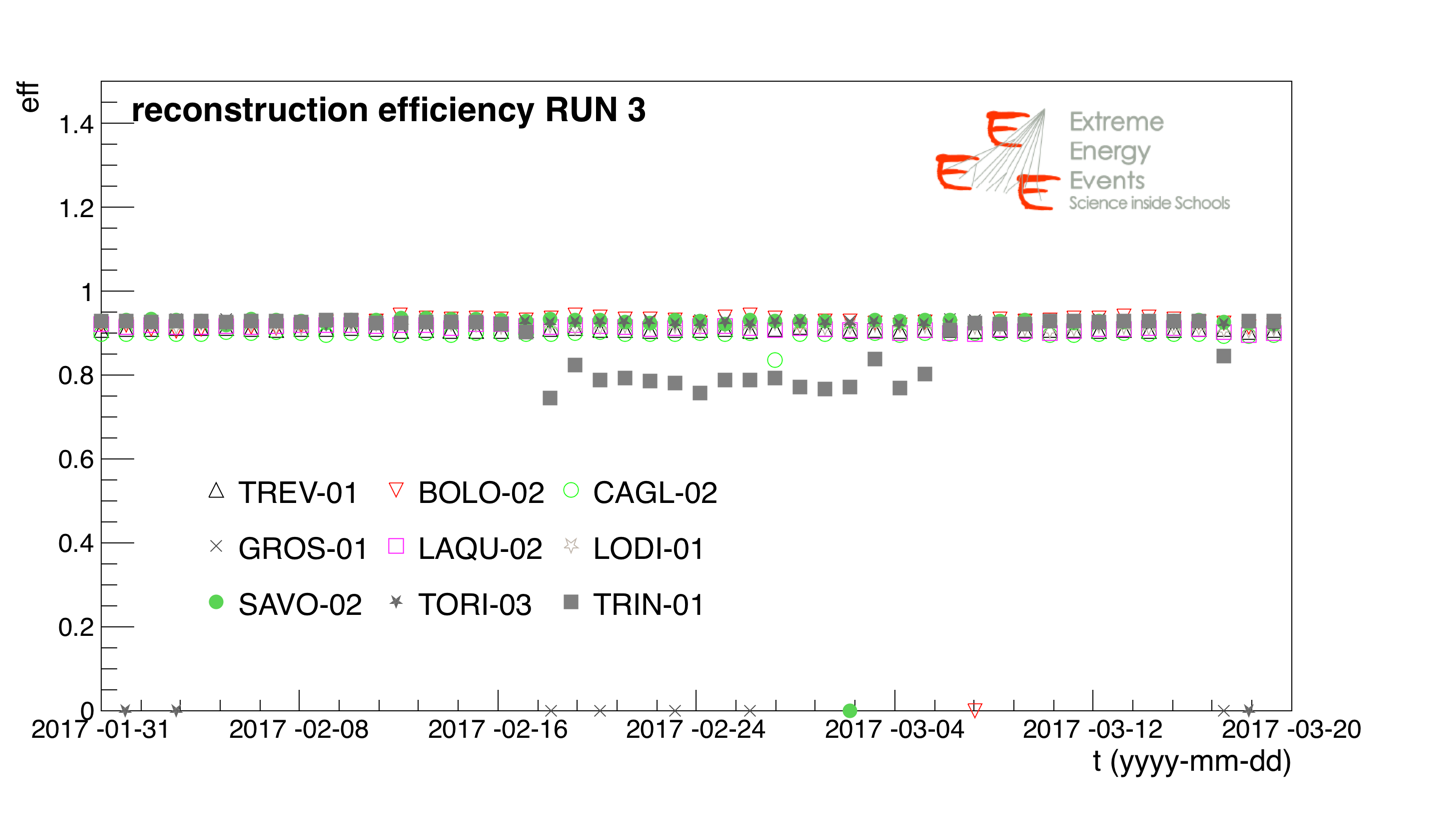}}  
	\hspace{0mm}
	\subfloat[Time of Flight]{
  		\includegraphics[width=0.5\columnwidth]
       {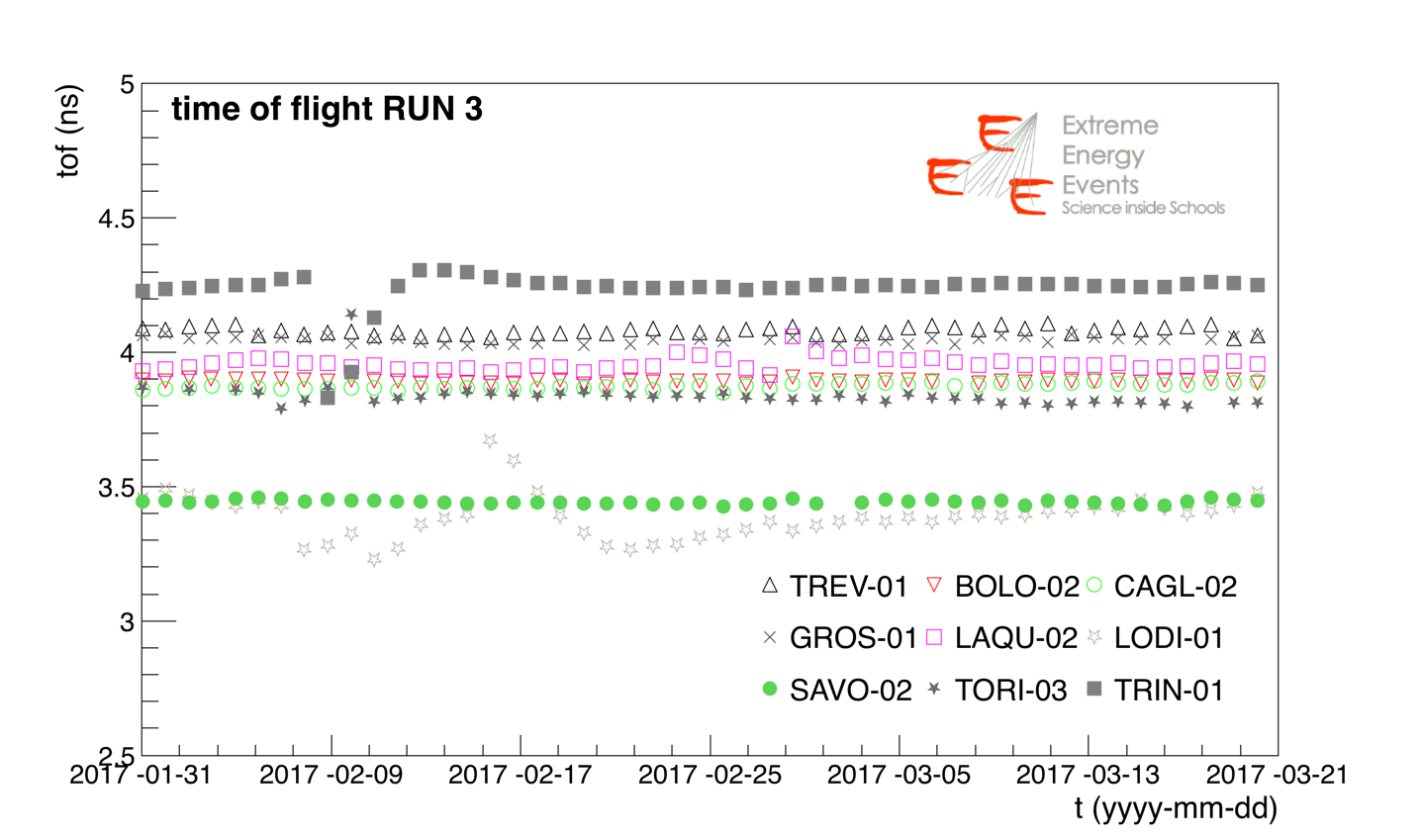}}
	\subfloat[Track rate]{
  		\includegraphics[width=0.5\columnwidth]
        {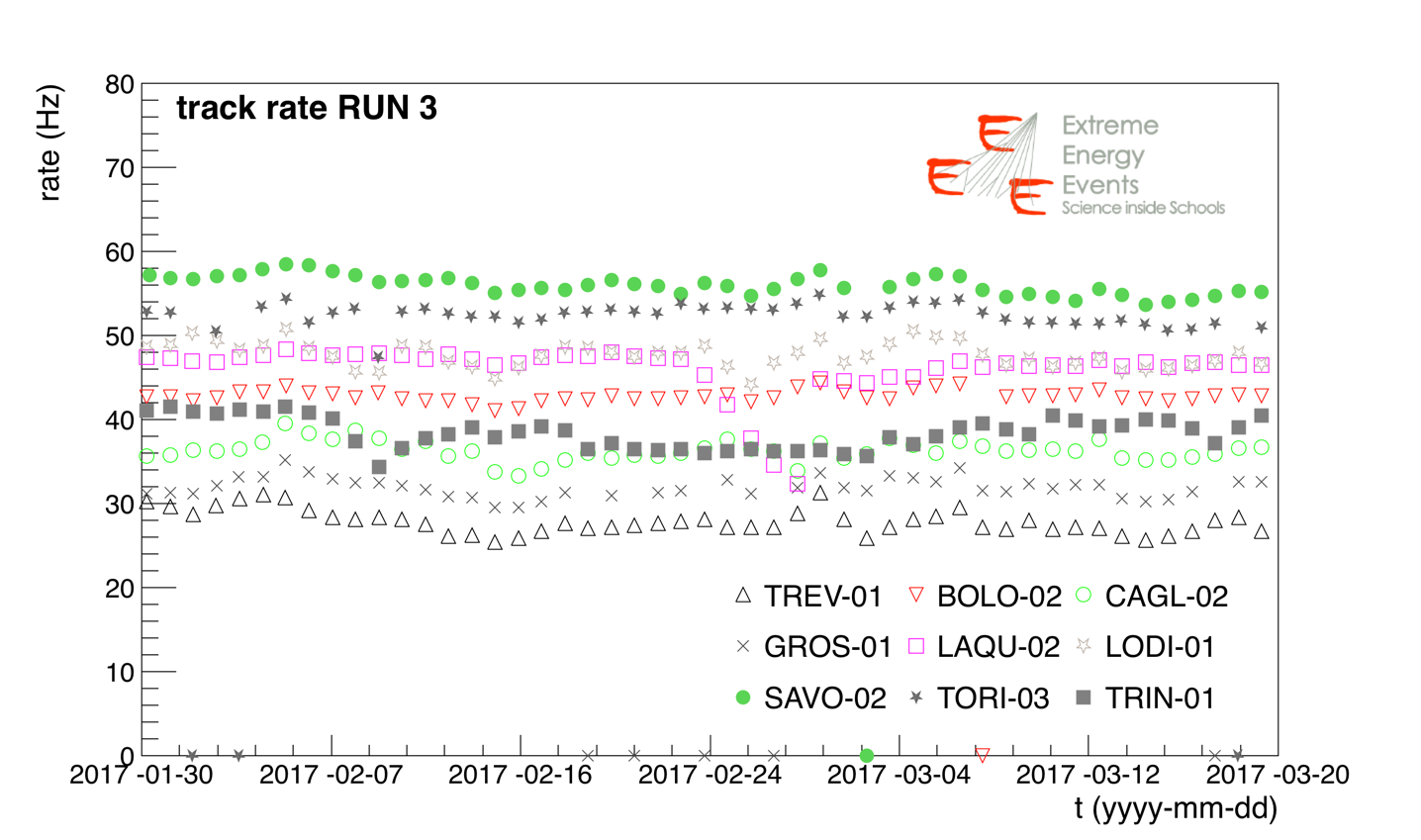}}
	\caption{\footnotesize Run 3 trending plots for 9 stations. Points at zero represent days when the telescope was not operational.}
	\label{Run3dataset}
\end{figure}
Data sample used in this case roughly covers 4.5 months of Run 2 for the the five EEE telescopes with the best ``live" time (time of active data taking). $\chi^2$ and TOF together can be used to check the quality of the reconstructed tracks. Mean TOF values may be different for each station since distance between chambers are not the same for all installations. $\chi^2$ is computed from the best (lowest $\chi^2$) tracks in each event, if at least one hit on each chamber has been recorded. Note that these rates are sensitive to fluctuations in efficiency and noise rate of the detector, with the reconstruction efficiency used as cross-check. Hit multiplicity is a key parameter, because is extremely sensitive to gas and HV instabilities. Hit multiplicity is also available in the DQM for each chamber independently, and the average and the single chamber multiplicity for some of the telescopes of Fig. \ref{Run2dataset} are shown in Fig. \ref{Run2Mult}.
\\Deviations from the standard trending provides important warnings for the identification of problems and their solution. They might be due to cables not correctly plugged-in, malfunctioning of parts of the electronics, gas bottle exhausted.
A long term stability study has been performed on 2 months out of Run 3 dataset and reported in Fig. \ref{Run3dataset}, showing the same quantities of Fig. \ref{Run2dataset} for a subset of 9 stations.
These plots are also useful to identify issues affecting a specific telescope for a limited period of time.
For instance correlation between sudden multiplicity increases and drops in track reconstruction efficiency and track rate is clearly evident.  

\section{Conclusions}

The network of cosmic muon telescopes of the EEE Project, based on MRPC technology and covering about 10$^{6}$ km$^{2}$ across the Italian territory, has been successfully operated in the last years.
More than 50 billion tracks have been collected by the network, during three data taking from 2014 to 2017.
The observatory has grown up by a factor almost 8 in terms of number of telescopes wrt. 2007 and the EEE network is currently the largest and long-living MRPC-based telescopes network, with 53 active sites and more than 12 years of data taking. The unconventional working sites offer a unique check of the robustness, the ageing  and the long-lasting performance of the MRPC technology for particle tracking and timing determination. 
The results of the analysis on the performance of the network are fully compatible with the EEE requirements in terms of efficiency ($\sim$93\%), time resolution (238 ps) and spatial resolution (1.5 cm and 0.9 cm respectively for longitudinal and transverse direction).
The good performance of the network allowed several analysis to be performed and published. Among them:
search coincidences between near telescopes \cite{coinc1}, study the muon flux decrease due solar events \cite{forbushDec2}\cite{forbushDec3}, study of cosmic muon anisotropy at sub-TeV scale \cite{anisotropies}, study of muon decay into up-going events \cite{upward}, search long distance correlations between EAS \cite{coinc2}.
At the moment the EEE Collaboration is focusing on further improvements of the performance in terms of duty cycle and optimization of the working points of the telescopes.


\end{document}